\documentclass[12pt,a4paper]{article}
\usepackage{amsmath,amsthm,amsfonts,amssymb,bbm}
\usepackage{graphicx,psfrag}
\usepackage{cite}

\numberwithin{equation}{section}
\renewcommand{\Re}{\mathrm{Re}}
\renewcommand{\Im}{\mathrm{Im}}

\newcommand{\Or}{\mathcal{O}}

\newcommand{\Pb}{\mathbbm{P}}
\newcommand{\E}{\mathbbm{E}}
\newcommand{\Id}{\mathbbm{1}}
\newcommand{\e}{\varepsilon}
\newcommand{\I}{{\rm i}}
\newcommand{\R}{\mathbb{R}}
\newcommand{\N}{\mathbb{N}}
\newcommand{\Z}{\mathbb{Z}}
\newcommand{\dx}{\mathrm{d}}
\newcommand{\Af}{{\cal A}_{\rm 1}}
\newcommand{\Ac}{{\cal A}_{\rm 2}}

\newtheorem{prop}{Proposition}
\newtheorem{thm}[prop]{Theorem}
\newtheorem{lem}[prop]{Lemma}
\newtheorem{defin}[prop]{Definition}
\newtheorem{cor}[prop]{Corollary}
\newtheorem{rem}[prop]{Remark}
\newenvironment{remark}{\begin{rem}\normalfont}{\end{rem}}

\newenvironment{proofOF}[2]{\removelastskip\vspace{6pt}\noindent {\it Proof of #1.}~\rm #2}{\qed \par\vspace{6pt}}

\title{Two speed TASEP}
\author{Alexei Borodin\thanks{California Institute of Technology,
e-mail: borodin@caltech.edu},
Patrik L. Ferrari\thanks{Bonn University,
e-mail: ferrari@uni-bonn.de},\\
Tomohiro Sasamoto\thanks{Chiba University, e-mail: sasamoto@math.s.chiba-u.ac.jp and Technische Universit\"at  M\"unchen, e-mail: sasamoto@ma.tum.de}}

\date{September 16, 2009}

\begin{document}
\maketitle \sloppy

\begin{abstract}
We consider the TASEP on $\Z$ with two blocks of particles having
different jump rates. We study the large time behavior of particles' positions.
It depends both on the jump rates and the region we focus on, and we
determine the complete process diagram. In particular, we
discover a new transition process in the region where the influence of
the random and deterministic parts of the initial condition interact.

Slow particles may create a shock, where the particle density is
discontinuous and the distribution of a particle's position is asymptotically
singular. We determine the diffusion coefficient of the shock
without using second class particles.

We also analyze the case where particles are effectively blocked by
a wall moving with speed equal to their intrinsic jump rate.
\end{abstract}

\section{Introduction} \label{SectIntro}
We consider the totally asymmetric simple exclusion process (TASEP)
on $\Z$. This is one of the basic one-dimensional interacting
stochastic particle systems that, despite its simplicity, is full of
interesting features. It consists of particles moving to the right
by jumps of length one. The jumps happen at a given rate (the clocks
of different particles are independent), and the particles are
subject to the exclusion constraint ---  no site can be occupied by more than one particle.
This model can also be seen as a growing interface with gradient
given by the particle density; it belongs to the KPZ
(Kardar-Parisi-Zhang) universality class of growth models.
Recently the fluctuation properties of the TASEP and related models
have been studied extensively using the techniques from random
matrix theory \cite{PS02,Sp05,FP06,Sas07}. See also
\cite{DG09,TW08a,TW08b} for more recent developments on
the case where particles can hop in both directions.

In previous works~\cite{Sas05,BFPS06,BFS07,BF07} we analyzed the
large time $t$ behavior of particles' positions for some non-random
initial conditions and uniform jump rate (say equal to one). For
example, if particles start from $2\Z_-$ then the large time
macroscopic density is given by
\begin{equation}
\varrho([\xi t],t)=\begin{cases}
1/2,&\textrm{if }\xi<0,\\
(1-\xi)/2,&\textrm{if }\xi\in [0,1],\\
0,&\textrm{if }\xi>1.
\end{cases}
\end{equation}
For large time $t$, the correlation length scales as $t^{2/3}$ and
the fluctuations scale as $t^{1/3}$. Under an appropriate scaling
limit, the joint distributions of particles' positions are governed
by a process which depends on the density gradient: (a) for $\xi<0$,
it is the Airy$_1$ process, (b) for $\xi\in(0,1)$ it is the Airy$_2$
process, and at $\xi=0$ it is the Airy$_{2\to 1}$ transition
process, see~\cite{BFS07}. Similarly, one can consider the joint
distributions of the current at different positions instead of the
positions of different particles --- the limit processes are
unchanged.

In this paper we consider a small variation of the above situation,
which however shows a number of  new nontrivial phenomena. Instead
of setting the jump rate to $1$ for all the particles, we modify the
jump rate of the first $M$ particles and set it equal to $\alpha>0$.

There are a few cases to consider. For example, for $0<\alpha<1/2$
the first (slow) particles generate a shock where the macroscopic
particle density changes discontinuously from $1/2$ to $1-\alpha$.
The fluctuations on the left of the shock are Airy$_1$-distributed
on the $t^{1/3}$~scale, while inside the jam region they are ${\rm
GUE}(M)$-distributed on the $t^{1/2}$ (diffusion) scale, see the
body of the paper for details\footnote{Here GUE($M$) stands for
the Gaussian Unitary Ensemble of $M\times M$ random matrices.}.
Also, the distribution of a particle's position in the shock region
has a singularity.

When $\alpha$ reaches $1/2$, the macroscopic density becomes
constant (it is equal to $1/2$ everywhere), but the fluctuations are
different. In the simplest case of $M=1$ slow particle, by Burke's
theorem~\cite{Bur56}, our initial condition is equivalent to the
deterministic one on $\Z_-$ (even sites are occupied as before) and
to the product of Bernoulli measures with density $1/2$ on $\Z_+$.
Thus, for $\alpha=1/2$ there is a transition region where the
influence of the initial randomness becomes relevant, but it can not
be seen macroscopically.

On Figure~\ref{FigDiagram} we present the whole \emph{process
diagram}\footnote{A bit like a phase diagram, but in our case
instead of phases and phase transition we have limit processes and
transition processes.}. One of the goals of the present paper is to
derive the large time fluctuations' behavior in all of its regions.

Another situation we consider is $M=\infty$ and $\alpha=2$. Under
$t^{1/2}$-scaling, the speed $\alpha$ particles effectively act as a
wall moving with speed $1$. The following $n$ ``normal'' particles
then become like Brownian motions with the first one being reflected
off the wall and the following ones being reflected off each other.
The large time fluctuations are then given by the
\emph{antisymmetric} ${\rm GUE}(M)$ process (for fixed time it was
characterized in~\cite{FN08}, see also \cite{D1}, \cite{D2}). This
is also closely related to the asymptotics of a certain Markovian
dynamics for two-dimensional interlacing particle systems with a
wall, see Section 2.3 of~\cite{WW08} and \cite{BK}. Using the
relation between last passage directed percolation with exp(1)
random variables and TASEP, one can predict that there should be a
relation between the maximum process for the largest eigenvalue of
the Dyson's Brownian Motion and systems of nonintersecting paths with
a wall. This relation will be made more precise in~\cite{BFPSW09}.

Our arguments are based on deriving suitable determinantal
expressions for the quantities of interest and analyzing the
resulting (Fredholm) determinants asymptotically. In most cases, a
mathematically rigorous argument of that kind would require the
evaluation of the asymptotics of the kernel under the determinant,
as well as some control over the decay of the kernel at infinity.
This last part is often viewed as a technicality, and we omit tail
estimates in the present paper.

In this determinantal approach, the main difficulty typically lies
in deriving an integral representation for the kernel before the
limit transition; evaluating the asymptotics is often quite
straightforward via the standard steepest descent analysis. However,
in the shock case mentioned above, we faced a new effect --- in the
large time limit the kernel diverged. We had been puzzled by this
difficulty for a while, and we view finding the modification of the
kernel that solved the problem as our main technical novelty.

\vspace{0.5em} \emph{Outline.} The rest of the paper is organized as
follows. In Section~\ref{SectModel} we explain the macroscopic
picture and describe the process diagram. Then we state the results
for the different parts of the diagram, which are proven in
Sections~\ref{SectJam}-\ref{SectNoSlow}.  In Section~\ref{SectKernels}
we obtain the determinantal correlation structure and the associated
kernel with a couple of specializations. Finally, in
Section~\ref{SectWall} we consider the reflecting wall situation.

\vspace{0.5em} \emph{Acknowledgments.} Borodin was partially supported by NSF grant DMS-0707163.
The work of Sasamoto was partially supported by the Grant-in-Aid for Young Scientists (B), the Ministry of Education, Culture, Sports, Science and Technology, Japan.

\section{Model and Results}\label{SectModel}
The continuous time TASEP on $\Z$ is a model of interacting particle
systems in which at every instant at most one particle occupy a site
in $\Z$. Particles jump by $1$ to the right with a given jump rate
provided the arriving site is empty. As a consequence, particles do
not overtake each other. Hence, we can assign labels to particles,
say particle $n$ has position at time $t$ equal to $x_n(t)$. We also
denote by $v_n$ the jump rate of particle $n$. Our convention is to
consider particles labeled from right to left, i.e.,
$x_n(t)>x_{n+1}(t)$ for any time $t$. We denote by $y_n=x_n(0)$ the
starting position of particle $n$ (non-random).

We always start with a finite number of particles, but since the interactions are due only to the blocking from the right, it is effectively equivalent to have the index $n$ varying over $\N$. Limiting cases when $n$ varies over $\Z$ can also be treated as appropriate limits of finite systems. For particle-dependent jump rates, we derived in~\cite{BF07} the general formula for the joint distribution of any subsets of particles at time $t$. This result is restated as Proposition~\ref{PropJointCorr}.

In order to apply this result, we need to set the initial positions $y_k$'s and the jump rates $v_k$'s. In this paper we consider the first $M$ particles to have jump rate $\alpha$ and the rest having unit jump rate:
\begin{equation}\label{eq3.8}
y_j=2(M-j),\qquad v_j=\left\{\begin{array}{ll}
\alpha, & 1\leq j \leq M,\\
1, & j>M.
\end{array}\right.
\end{equation}
The choice of setting the last $\alpha$-particle at the origin is due to a simplification in the specific situation where we will take the $M\to\infty$ limit.

\subsubsection*{Macroscopic description for $0<\alpha<1$}
The fluctuation results will depend on the macroscopic behavior, so let us first describe it. By macroscopic scale we mean when spatial directions are linearly scaled with time $t$. On that scale, for $\alpha\in (0,1)$, the effect of finitely many slow particle(s) is equivalent to having a starting density of rate $1$ particles on $\N$ equal to $1-\alpha$. A particularly important case is $M=1$, for which by Burke's Theorem~\cite{Bur56}, the initial condition is exactly equal to alternating deterministic on $\Z_-$ and Bernoulli-$(1-\alpha)$ on $\N$.

\begin{figure}[t]
\begin{center}
\psfrag{1}[rb]{$1$}
\psfrag{1-a}[rb]{$1-\alpha$}
\psfrag{1/2}[lb]{$1/2$}
\psfrag{a-1/2}[cc]{$(\alpha-1/2)t$}
\psfrag{a}[cc]{$\alpha t$}
\psfrag{mu}[rc]{$(2\alpha-1)t$}
\psfrag{rho}[lb]{$\rho(x,t)$}
\psfrag{x}[cb]{$x$}
\psfrag{(a)}[cb]{$(a)$}
\psfrag{(b)}[cb]{$(b)$}
\psfrag{t=0}[lb]{$t=0$}
\psfrag{t>>1}[lb]{$t\gg 1$}
\includegraphics[height=7cm]{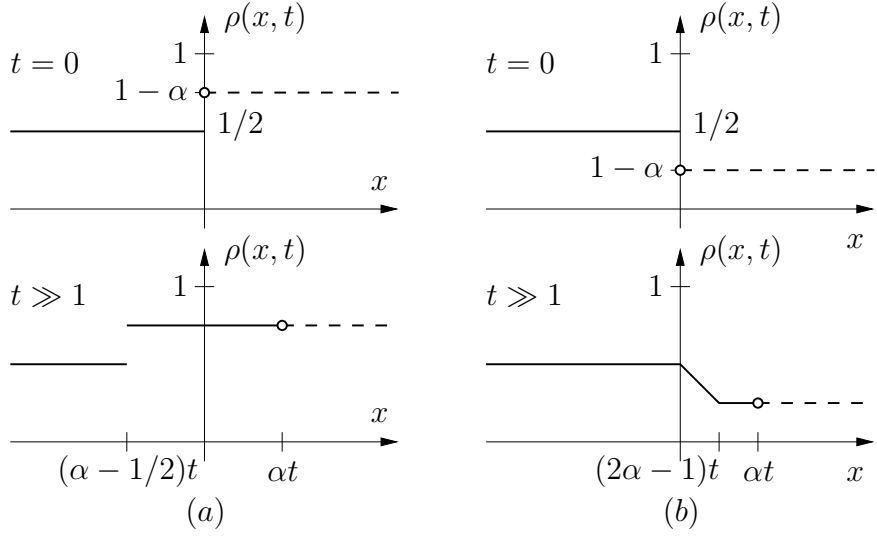}
\caption{Particles' density for large $t$: $(a)$ for $0<\alpha<1/2$, $(b)$ for $1/2<\alpha<1$. The big dot is the position of the right-most slow particle. }
\label{FigMacroscopic}
\end{center}
\end{figure}

Let $\varrho(\xi,\tau)$ be the macroscopic density of particles,
\begin{equation}
\varrho(\xi,\tau)=\lim_{t\to\infty}\Pb(\textrm{there is a particle at }[\xi t]\textrm{ at time }\tau t)
\end{equation}
Then, the average current from TASEP dynamics through position $[\xi t]$ at time $\tau t$ is given by $\varrho(1-\varrho)$, from which it follows that $\varrho$ satisfies the Burgers' equation~\cite{Rez91}
\begin{equation}\label{eqBurgers}
\partial_\tau \varrho+\partial_\xi (\varrho(1-\varrho))=0.
\end{equation}
To get the large-time macroscopic density one has to solve (\ref{eqBurgers}) with initial condition
\begin{equation}\label{eqMacroIC}
\varrho(\xi,0)=\left\{
\begin{array}{ll}
1/2,&\textrm{for }\xi<0,\\
1-\alpha,&\textrm{for }\xi\geq 0.
\end{array}\right.
\end{equation}
The solution at $\tau=1$ is as follows. For $\alpha\in [0,1/2)$, it has a discontinuity at $\xi=\alpha-1/2$ \, (in this case, one needs to use a conservation law to obtain this solution, see e.g.~\cite{Whi99}),
\begin{equation}
\varrho(\xi,1)=\left\{\begin{array}{ll}
1/2,&\textrm{if }\xi<(\alpha-1/2),\\
1-\alpha,&\textrm{if }\xi>(\alpha-1/2),
\end{array}\right.
\end{equation}
while for $\alpha\in [1/2,1]$
\begin{equation}
\varrho(\xi,1)=\left\{\begin{array}{ll}
1/2,&\textrm{if }\xi\leq 0,\\
(1-\xi)/2,&\textrm{if }\xi\in [0,2\alpha-1],\\
1-\alpha,&\textrm{if }\xi\geq 2\alpha-1.
\end{array}\right.
\end{equation}
So, for large $t$, the density of particles in the lattice-scale,
\begin{equation}
\rho(x,t)=\Pb(\textrm{there is a particle at }x\textrm{ at time }t)\cong \varrho(x/t,1),
\end{equation}
see Figure~\ref{FigMacroscopic} for an illustration.

A consequence is that the number of particles moving with average speed $\alpha$ is around $(1-\alpha)t/2$ for $\alpha\leq 1/2$ and $(1-\alpha)^2t$ for $1/2\leq \alpha<1$. Also, the macroscopic position at time $t$ of particle $n=[\nu t]$ is
\begin{equation}\label{eqMacroX}
\mathbf{x}_{\alpha}(\nu ):=\lim_{t\to\infty} t^{-1}\E(x_{[\nu t]}(t))= \left\{\begin{array}{ll}
\alpha-\nu/(1-\alpha),&\textrm{if }\nu \in (0,\min\{\tfrac{1-\alpha}{2},(1-\alpha)^2\}),\\
1-2\sqrt{\nu},&\textrm{if }\nu \in((1-\alpha)^2,\tfrac14),\\
1/2-2\nu,&\textrm{if }\nu >\max\{\tfrac{1-\alpha}{2},\tfrac14\},
\end{array}\right.
\end{equation}
where the second case occurs only for $\alpha\in (1/2,1]$.

\subsubsection*{Process diagram}
\begin{figure}
\begin{center}
\psfrag{0}[cb]{$0$}
\psfrag{14}[cb]{$1/4$}
\psfrag{12}[cb]{$1/2$}
\psfrag{1}[cb]{$1$}
\psfrag{b}[lb]{$n/t$}
\psfrag{a}[cb]{$\alpha$}
\psfrag{A1}[cb]{$\Af$}
\psfrag{A2}[cb]{$\Ac$}
\psfrag{OU}[lb]{${\rm DBM}$}
\includegraphics[height=5cm]{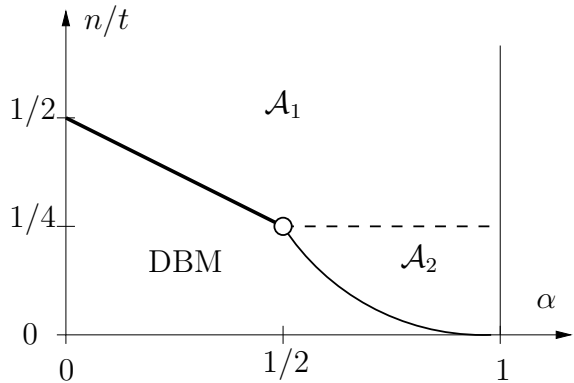}
\caption{Process diagram. The black thick line is the shock. On the dashed line there is the Airy$_{2\to 1}$ process. At the white dot there is the Airy$_{2\to 1,M,\kappa}$ process. At the curved solid line the process is the Airy$_{{\rm DBM}\to 2}$.}
\label{FigDiagram}
\end{center}
\end{figure}

As we have seen, there are different types of macroscopic behavior for the density. For example, when $\alpha\in (1/2,1)$, there are two plateaux in the density joined by a linearly decreasing part. The plateaux are of different nature, since only the right one is influenced by the $\alpha$-particles. So, the limit process of particles' position varies depending on which part of the process diagram the parameter are in, see Figure~\ref{FigDiagram}.

For keeping the presentation of the results as simple as possible, the results stated in the remainder of this section are the particularization to fixed time. However, the results hold in greater generality and span from the fixed time to tagged particle problem. The general statements are contained in the following sections.

\vspace{0.5em}
\noindent\emph{(1) Dyson's Brownian Motion region.} For fixed time $t$ let us consider particles with number $n=[\nu t]$ with $\nu \in(0,\min\{\tfrac{1-\alpha}{2},(1-\alpha)^2\})$, i.e., we are in the right plateau with density $1-\alpha$. In the diffusion scaling limit, the $M$th $\alpha$-particle has ${\rm GUE}(M)$ distributed fluctuations. So, to get a non-trivial limit for the particles in the jammed region we have to look at fluctuations with respect to the macroscopic behavior on the $t^{1/2}$~scale. Therefore, we set the rescaled process as
\begin{equation}\label{eqA}
X_t(\nu):=\frac{x_{[M+\nu t]}(t)-\mathbf{x}_{\alpha}(\nu) t}{-\sigma(\nu) t^{1/2}},\quad \sigma^2(\nu)=\alpha(1-\nu/(1-\alpha)^2),
\end{equation}
and $\mathbf{x}_{\alpha}(\nu)=\alpha+\nu/(1-\alpha)$, see~(\ref{eqMacroX}). Then
\begin{equation}
\lim_{t\to\infty} X_t(\nu)={\rm DBM}(-\ln\sigma(\nu)).
\end{equation}
${\rm DBM}$ is the stationary process of eigenvalues of $\beta=2$ Dyson's Brownian Motion on $M\times M$ Hermitian matrices, see Lemma~\ref{OU_kernel} for a definition. The change in time $-\ln\sigma(\nu)$ is simply due to the non-stationarity of $X_t(\nu)$, while ${\rm DBM}$ is stationary. The complete statement is in Proposition~\ref{PropJam}.

\begin{figure}[t]
\begin{center}
\psfrag{x}[c]{$x$}
\psfrag{n}[l]{$n$}
\psfrag{t12}[cb]{$t^{1/2}$}
\psfrag{xshock}[c]{$x_{\rm shock}$}
\psfrag{nshock}[l]{$n_{\rm shock}$}
\includegraphics[height=5cm]{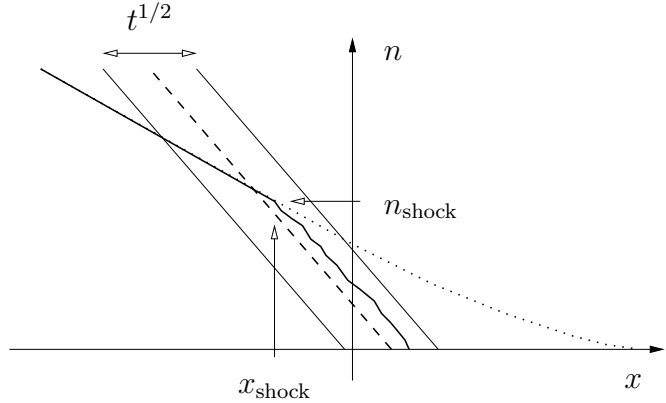}
\caption{Illustration of the shock. The continuous line is the position of particles, the dashed line is the macroscopic position of particles inside the jam, while the dotted line would be the position without the $\alpha$-particle.  See~\cite{FerTASEP} for an animation of the TASEP with and without a slow particle.}
\label{FigureShock}
\end{center}
\end{figure}

\vspace{0.5em}
\noindent\emph{(2) Shock region, $M=1$.} Consider now $\alpha\in (0,1/2)$, where there is a macroscopic shock traveling to the left with speed $(\alpha-1/2)$, and consider the important case of Bernoulli-$(1-\alpha)$ on $\Z_+$ as initial condition, that is $M=1$. So, if a particle is already inside the jam, then it has $t^{1/2}$ fluctuations with respect to the dashed line in Figure~\ref{FigureShock}. On the other hand, particles can not move faster than they would in absence of the $\alpha$-particles, in which case they fluctuate on a $t^{1/3}$~scale around the dotted line in Figure~\ref{FigureShock}. So, on the $t^{1/2}$~scale, the dotted line acts as a sharp cut-off and the result is the following.

\begin{prop}\label{PropShock}
Consider one slow particle with $0<\alpha<1/2$ and the scaling
\begin{equation}\label{eqShock}
\begin{aligned}
n&=
\tfrac{1-\alpha}{2}t+\eta t^{1/2},\\
x(\xi)&=
\tfrac{1}{2}t-2 n-\xi t^{1/2}.
\end{aligned}
\end{equation}
For $\xi>0$,
\begin{equation}\label{eqShock2}
\lim_{t\to\infty} \Pb(x_n(t)\geq x(\xi))=\frac{1}{\sqrt{2\pi\sigma^2}}\int_{-\infty}^{\xi+\xi_c} \dx y \exp(-y^2/2\sigma^2),
\end{equation}
where
\begin{equation}
\sigma^2=\frac{\alpha(1-2\alpha)}{2(1-\alpha)},\quad \xi_c=\frac{1-2\alpha}{1-\alpha}\eta.
\end{equation}
For $\xi<0$:
\begin{equation}\label{eqShock3}
\lim_{t\to\infty} \Pb(x_n(t)\geq x(\xi))=0.
\end{equation}
\end{prop}
Geometrically, $\xi=0$ are points on the dotted line, while $\xi=\xi_c$ are on the dashed line of Figure~\ref{FigureShock}. This proposition is proved in Section~\ref{subsectShock}.

More precisely the picture is as follows: On the $t^{1/2}$~scale,
the random interface given by $\{(x_n(t),n),n\geq 1\}$ looks like a
plot of a (non-homogeneous) Ornstein-Uhlenbeck process with average
given by the dashed line. When the trajectory hits the dotted line,
it sticks to it and does not fluctuate anymore. We identify this
hitting position as the shock. This picture is consistent with
Proposition~\ref{PropShock}. A consequence of
Proposition~\ref{PropShock} is that the shock position has Gaussian
fluctuations (for $M=1$) with diffusion coefficient
$D=\frac{\alpha(1-\alpha)}{1/2-\alpha}$. This is the content of
Proposition~\ref{PropShockDiffusionCoeff}. The main novelty in this
result is that we do not start with Bernoulli initial conditions on
both $\Z_-$ and $\Z_+$ with two different densities and, more
interestingly, we do not have to introduce second class particles to
define the shock position, as it was the case for example
in~\cite{Fer90,DJLS93,CFP08}.

\vspace{0.5em}
\noindent\emph{(3) Transitions and Airy processes.} The last two new results related to the process diagram are the transition point (the white dot in Figure~\ref{FigDiagram}) and the transition line between ${\rm DBM}$ and the Airy$_2$ process (along the curved line in Figure~\ref{FigDiagram}). The other remaining regions, are the ones where the influence of the $\alpha$-particles is not present, so one clearly gets the same results obtained in~\cite{BFS07}.

The limit processes are different, but the scaling limit can be presented in the same way for all the cases.
Indeed, consider $n\sim \nu t$ with
\begin{equation}
\begin{aligned}
\nu &>(1-\alpha)/2,&\textrm{for }\alpha\in (0,1/2),\\
\nu &\geq (1-\alpha)^2,&\textrm{for }\alpha\in [1/2,1).
\end{aligned}
\end{equation}
Then, the rescaled process for fixed time $t$ is given by
\begin{equation}
X_t(\tau):=\frac{x_{[M+\nu t-2\tau t^{2/3}]}(t)-\mathbf{x}_{\alpha}(\nu-2\tau t^{-1/3}) t}{-t^{1/3}}.
\end{equation}

The new transition process is at the macroscopic point given by $\nu=1/4$ and with $\alpha=\frac12(1+\kappa t^{-1/3})$. In Theorem~\ref{ThmTransition} of Section~\ref{SectTransition} we prove that
\begin{equation}
\lim_{t\to\infty} X_t(\tau)=S_v{\cal A}_{2\to 1,M,\kappa S_v}(\tau/S_h)
\end{equation}
with $S_v=2^{-1/3}$, $S_h=2^{-5/3}$, and $\tau\mapsto{\cal A}_{2\to 1,M,\kappa}(\tau)$ is given in Definition~\ref{DefTransition}.

The second transition process is at the line $\nu =(1-\alpha)^2$, for \mbox{$\alpha\in (1/2,1)$}. In Theorem~\ref{ThmTransitionToGUE} of Section~\ref{SectTransition} we prove that
\begin{equation}
\lim_{t\to\infty} X_t(\tau)=S_v{\cal A}_{{\rm DBM}\to 2}(\tau/S_h)
\end{equation}
with $S_v=\left(\frac{(1-a)\alpha}{(1-\alpha)(2-\alpha)}\right)^{1/3}$, $S_h=\frac{(1-\alpha)^2}{\alpha} S_v^2$, and $\tau\mapsto{\cal A}_{{\rm DBM}\to 2}(\tau)$ is given in Definition~\ref{DefTransitionGUE}. The process ${\cal A}_{{\rm DBM}\to 2}(\tau)$ has appeared before, see~\cite{BBP04,SI07,BP07,FAvM08} (with sometimes the time direction inverted).

Finally, to complete the process diagram we state what the limiting processes are in the $\alpha$-independent cases in Section~\ref{SectNoSlow}, where one has either the Airy$_1$, the Airy$_2$ or the Airy$_{2\to 1}$ process. The fixed time $t$ results were already contained in~\cite{BFP06}.

\subsubsection*{TASEP with a reflecting wall}
The last result of this paper is of a different nature, since we do not have slow particles. Instead, consider the case $\alpha=2$ but with $M=\infty$. Then the particle that started at the origin moves with average speed $1$ and has fluctuations of order $t^{1/3}$. The other ``normal'' particles, have jump rate $1$ and are blocked by the last $\alpha$-particle, but in contrast to the jam situation they do not have the tendency of filling up the gap rapidly, since the ``wall'' moves with their natural speed. Let us call particle $n$ the one starting at $-2n$. Particle $1$ fluctuates on a $t^{1/2}$~scale, that means that from its perspective the last $\alpha$-particle is like a moving blocking wall. Viewed from the ``wall'', particle $1$ does essentially a reflected random walk in continuous time, and particle $2$ does a random walk reflected on particle $1$ and so on.

Consider a sequence of particle numbers $n_i$ (not rescaled with time) and times $t_i=\tau_i t$. Since particles at time $t_i$ are approximately at position $t_i$ (speed one), we define the rescaled random variables
\begin{equation}
i\mapsto X_t(i)=\frac{x_{n_i}(t_i)-t_i}{-\sqrt{2t_i}}.
\end{equation}

We can compute the correlation functions of $X_t(i)$'s only if they are \emph{space-like}, property denoted by $\sim$ and defined by
\begin{equation}
(n_1,t_1)\sim (n_2,t_2) \iff (n_1,t_1)\prec (n_2,t_2)\textrm{ or }(n_2,t_2)\prec (n_1,t_1)
\end{equation}
with
\begin{equation}
(n_1,t_1)\prec (n_2,t_2) \Leftrightarrow n_1\leq n_2,t_1\geq t_2,\textrm{ and are not identical}.
\end{equation}

Then, our result proven in Section~\ref{SectWall} is the following.
\begin{thm}\label{thmWall}
For any given $m=1,2,\ldots$, let us choose $m$ space-like couples $(n_i,\tau_i)$, $1\leq i\leq m$. Let $\rho^{(m)}_t(\xi_1,\ldots,\xi_m)$ be the $m$-point correlation functions of $X_t(1),\ldots,X_t(m)$. Then
\begin{equation}
\lim_{t\to\infty} \rho^{(m)}_t(\xi_1,\ldots,\xi_m) = \det\left[K^{\rm aGUE}((n_i,\theta_i),\xi_i;(n_j,\theta_j),\xi_j)\right]_{1\leq i,j\leq m}
\end{equation}
where $\theta_j=\ln(\tau_i)$.
\end{thm}
The kernel $K^{\rm aGUE}$ is an extension of the antisymmetric GUE minor kernel~\cite{FN08} defined as follows (see Lemma~\ref{LemmaIntReprAsymGUE} for an integral representation).
\begin{defin}\label{DefaGUE}
The extended kernel $K^{{\rm aGUE}}$ is defined by
\begin{multline}\label{aGUEmK}
K^{{\rm aGUE}}((n_1,\theta_1),\xi_1;(n_2,\theta_2),\xi_2)\\ =
\frac{2}{\sqrt{\pi}}e^{-\xi_1^2}
\sum_{\ell\in I}\frac{{\rm sign}(\ell)e^{-(\theta_2-\theta_1)\ell}}{2^{n_2+1-2\ell}(n_2+1-2\ell)!} H_{n_1+1-2\ell}(\xi_1) H_{n_2+1-2\ell}(\xi_2)
\end{multline}
where ${\rm sign}(\ell):=1$ if $\ell\geq 1$, ${\rm sign}(\ell):=-1$ if $l\leq 0$, and the interval of summation $I$ is
\begin{equation}
\begin{aligned}
I&=\{1,2\ldots,\lfloor (n_2+1)/2 \rfloor\}, &\textrm{if }(n_1,\theta_1) \not\prec (n_2,\theta_2), \\
I&=\{-\infty,\ldots,-1,0\}, &\textrm{if }(n_1,\theta_1) \prec (n_2,\theta_2).
\end{aligned}
\end{equation}
The functions $H_k(x)$ are the standard Hermite polynomials with normalization
$\int_\R\dx x H_k(x) H_j(x) e^{-x^2}=\delta_{k,j} k! 2^k \sqrt{\pi}$.
\end{defin}

In the fixed-$n$ specialization of the $K^{{\rm aGUE}}$ kernel, i.e., for $n_1=n_2=n$, this kernel becomes the one of a system of $N$ non-intersecting Brownian motions (rescaled to become stationary) as follows:\\[0.5em]
(a) if $n=2N-1$, the Brownian motions have the reflecting wall at the origin,\\[0.5em]
(b) if $n=2N$, the Brownian motions have the absorbing wall at the origin.\\[0.5em]
These processes were introduced and studied in~\cite{KT04,KT07} but the kernels were
not explicitly provided. Also, the kernel for $n=2N$ was obtained in the
study of $N$ Brownian excursions~\cite{TW07b}.

\section{Determinantal structure and kernels}\label{SectKernels}
We start by stating the general formula and then particularize to our choice of jump rates. Consider particles numbered by $1,2,\ldots$, with particle $j$ starting from site $y_j$ and jumping to the right with hopping rate $v_j$. The joint distributions of particle positions are obtained as a specialization $a(t)=t,b(t)=0$ of Proposition 3.1 in~\cite{BF07}. To state the result, consider the set of numbers $\{v_1,\ldots,v_n\}$ and let \mbox{$\{u_1<u_2<\ldots<u_{\nu}\}$} be their different values, with $\alpha_k$ being the multiplicity of $u_k$. Then we define a space of functions in $x$,
\begin{equation}
V_n=\mathrm{span}\{x^l u_k^x ,1\leq k \leq \nu, 0\leq l \leq \alpha_k-1\}.
\label{eqVn}
\end{equation}
The next statement holds for finite sequences of (distinct) events in the $(n,t)$ variables which are \emph{space-like}.
\begin{prop}\label{PropJointCorr}
Let us consider particles starting from $y_1>y_2>\ldots$ and denote
$x_j(t)$ the position of $j$th particle at time $t$. Take a sequence
of particles and times which are space-like, i.e., a sequence of $m$ couples
${\cal S}=\{(n_k,t_k),k=1,\ldots,m\,|\, (n_k,t_k)\prec (n_{k+1},t_{k+1})\}$.
The joint distribution of their positions $x_{n_k}(t_k)$ is given by
\begin{equation}\label{joint}
\Pb\Big(\bigcap_{k=1}^m \big\{x_{n_k}(t_k) \geq a_k\big\}\Big)=
\det(\Id-\chi_a K\chi_a)_{\ell^2(\{(n_1,t_1),\ldots,(n_m,t_m)\}\times\Z)}
\end{equation}
where $\chi_a((n_k,t_k),x)=\Id(x<a_k)$. Here $K$ is the kernel with
entries
\begin{equation}
\label{eqKernelFinal}
K((n_1,t_1),x_1;(n_2,t_2),x_2) = -\phi^{((n_1,t_1),(n_2,t_2))}(x_1,x_2)+\overline{K}((n_1,t_1),x_1;(n_2,t_2),x_2)
\end{equation}
where
\begin{equation}
\begin{aligned}\label{eqK0}
&\overline{K}((n_1,t_1),x_1;(n_2,t_2),x_2)
=\sum_{k=1}^{n_2} \Psi^{n_1,t_1}_{n_1-k}(x_1) \Phi^{n_2,t_2}_{n_2-k}(x_2),\\
&\phi^{((n_1,t_1),(n_2,t_2))}(x_1,x_2)
=\frac{1}{2\pi\I}\oint_{\Gamma_{0,\vec{v}}}\frac{\dx w}{w}\frac{e^{(t_1-t_2)w}}{w^{x_1+n_1-x_2-n_2}}\frac{\Id_{[(n_1,t_1)\prec (n_2,t_2)]}}{(w-v_{n_1+1})\cdots (w-v_{n_2})}.
\end{aligned}
\end{equation}
With $\vec{v}$ we mean $\{v_{n_1+1},\ldots,v_{n_2}\}$. The contour $\Gamma_{0,\vec{v}}$ is any anticlockwise oriented loop that includes $0$ and the elements of $\vec{v}$. The functions $\Psi^{n,t}_{n-j}$, $j\geq 1$ are given by
\begin{equation}\label{eqPsi}
 \Psi_{n-j}^{n,t}(x) = \frac{1}{2\pi\I}\oint_{\Gamma_{0,\vec{v}}}\frac{\dx w}{w} \frac{e^{tw}}{w^{x-y_{j}+n-j}}\frac{\prod_{k=1}^n(w-v_{k})}{\prod_{k=1}^j(w-v_{j})}.
\end{equation}
The functions
$\{\Phi^{n,t}_{n-j}\}_{1\leq j\leq n}$ are characterized by the two
conditions:
\begin{equation}\label{ortho}
\langle \Phi^{n,t}_{n-j},\Psi^{n,t}_{n-k}\rangle:= \sum_{x\in\Z}\Phi^{n,t}_{n-j}(x) \Psi^{n,t}_{n-k}(x) = \delta_{j,k},
\quad 1\leq j,k\leq n,
\end{equation}
and ${\rm span}\{\Phi^{n,t}_{n-j}(x),1\leq j\leq n\}=V_n$.
\end{prop}

The notation $\frac{1}{2\pi\I}\oint_{\Gamma_K}\dx z f(z)$ here and below means that the integration path $\Gamma_K$ goes around the poles of $f(z)$ which are in the set $K$.

In our situation, the orthogonalization gives the following result.
\begin{lem}\label{LemmaPsiPhi}
For $n\leq M$, the $n$ orthogonal functions are
\begin{equation}
\begin{aligned}
 \Psi_{n-j}^{n,t}(x) &= \frac{1}{2\pi \I}\oint_{\Gamma_{0,1}} \frac{\dx w}{w} \frac{(w(w-\alpha))^{n-j} e^{tw}}{w^{x+2n-2M}},  \\
 \Phi_{n-j}^{n,t}(x) &= \frac{1}{2\pi \I} \oint_{\Gamma_{\alpha-1}}\dx v  \frac{(1+v)^{x+2n-2M}}{e^{t(v+1)}((v+1)(v+1-\alpha))^{n-j+1}}(2v+2-\alpha),
\end{aligned}
\end{equation}
where $j=1,\ldots,n$. For $n\geq M+1$, we have two cases:\\
(a) for $j=M+1,\ldots,n$,
\begin{equation}
\begin{aligned}\label{eq3.10}
 \Psi_{n-j}^{n,t}(x) &= \frac{1}{2\pi \I}\oint_{\Gamma_{0,1}} \frac{\dx w}{w} \frac{(w(w-1))^{n-j} e^{tw}}{w^{x+2n-2M}}, \\
 \Phi_{n-j}^{n,t}(x) &= \frac{1}{2\pi \I}\oint_{\Gamma_0} \dx v \frac{(1+v)^{x+2n-2M}}{e^{t(v+1)}(v(1+v))^{n-j+1}}(1+2v),
\end{aligned}
\end{equation}
(b) for $j=1,\ldots,M$,
\begin{equation}
\begin{aligned}\label{eq3.14}
 \Psi_{n-j}^{n,t}(x) &= \frac{1}{2\pi \I}\oint_{\Gamma_{0,\alpha}} \frac{\dx w}{w} \frac{(w(w-1))^{n-M} (w(w-\alpha))^{M-j} e^{tw}}{w^{x+2n-2M}}, \\
 \Phi_{n-j}^{n,t}(x) &= \frac{1}{(2\pi \I)^2} \oint_{\Gamma_{\alpha-1}}\dx v \oint_{\Gamma_{0,v}}\dx z \frac{(1+2z)(2v+2-\alpha)}{(z-v)(z+v+1)}\\
 &\times \frac{(1+z)^{x+2n-2M}}{e^{t(z+1)}(z(1+z))^{n-M}}\frac{1}{((v+1)(v+1-\alpha))^{M-j+1}}.
\end{aligned}
\end{equation}
\end{lem}

Our original derivation of the orthogonal functions was based on the known orthogonal functions for the case where all the particles have the same jump rate (see~\cite{Sas05,BFPS06}), and then by employing Gram-Schmidt orthogonalization procedure. A similar procedure could be used also for more than one jump rate different from one, but we did not do it. However, once the orthogonal functions are determined, it is easier to verify the orthogonality by direct computation, and this is what we do below.
\begin{proofOF}{Lemma~\ref{LemmaPsiPhi}}
The formulas for $\Psi^{n,t}_{n-j}(x)$ is just a simple substitution of (\ref{eq3.8}) into (\ref{eqPsi}). Notice that for $x<2M-2n$, $\Psi^{n,t}_{n-j}(x)=0$. Therefore,
\begin{equation}
\sum_{x\in\Z}\Phi^{n,t}_{n-j}(x)\Psi^{n,t}_{n-k}(x)=\sum_{x\geq 2M-2n}\Phi^{n,t}_{n-j}(x)\Psi^{n,t}_{n-k}(x),
\end{equation}
in which $x$-dependent terms under the integrals are given by
\begin{equation}
\sum_{x\geq 2M-2n}\left(\frac{1+v}{w}\right)^{x+2n-2M} = \frac{w}{w-(1+v)}\quad\textrm{provided} \quad|w|>|1+v|.
\end{equation}
With these preparations we can prove the orthogonal relation needed by the theorem.\\[1em]
\emph{Case $n\leq M$:} We have
\begin{equation}
\begin{aligned}
&\langle \Phi^{n,t}_{n-j},\Psi^{n,t}_{n-k}\rangle \\
=& \frac{1}{(2\pi\I)^2}\oint_{\Gamma_{\alpha-1}} \dx v \oint_{\Gamma_{0,1+v}}\dx w \frac{(w(w-\alpha))^{n-k} e^{tw}(2v+2-\alpha)}{((v+1)(v+1-\alpha))^{n-j+1} e^{t(v+1)}}\frac{1}{w-(v+1)} \\
=&\frac{1}{2\pi\I}\oint_{\Gamma_{\alpha-1}}\dx v (2v+2-\alpha) ((v+1)(v+1-\alpha))^{j-k-1} \\
=&\frac{1}{2\pi\I}\oint_{\Gamma_0}\dx z z^{j-k-1}=\delta_{k,j},
\end{aligned}
\end{equation}
where we used the fact that after summing over $x$ the pole at $w=0$ disappeared (since $n-j\geq 0$) and then the change of variable $z=(v+1)(v+1-\alpha)$ with $\dx z = (2v+2-\alpha)\dx v$.\\[1em]
\indent\emph{Case $n\geq M+1$:} We have to do four computations, depending on whether $j,k$ are larger or smaller than $M$.\\[0.5em]
\indent\emph{Case $j,k\geq M+1$:} We get $\langle \Phi^{n,t}_{n-j},\Psi^{n,t}_{n-k}\rangle=\delta_{k,j}$ by the same computation as in the $n\leq M$ case but with $\alpha$ replaced by $1$.\\[0.5em]
\indent\emph{Case $j,k\leq M$:} Also in this case, after summing over $x$ the pole at $w=0$ disappears but instead there is a simple pole at $w=z+1$. This can be easily integrated out and we get
\begin{equation}
\begin{aligned}
\label{eq3.17}
& \langle \Phi^{n,t}_{n-j},\Psi^{n,t}_{n-k}\rangle\\
 &=\frac{1}{(2\pi\I)^2}\oint_{\Gamma_{\alpha-1}}\dx v \oint_{\Gamma_{0,v}} \dx z \frac{((z+1)(z+1-\alpha))^{M-k}}{((v+1)(v+1-\alpha))^{M-j+1}} \frac{(1+2z)(2v+2-\alpha)}{(z-v)(z+v+1)} \\
&=\frac{1}{2\pi \I}\oint_{\Gamma_{\alpha-1}}\dx v ((v+1)(v+1-\alpha))^{j-k-1} (2v+2-\alpha)=\delta_{j,k}
\end{aligned}
\end{equation}
where we used that after integrating out the $w=z+1$ pole, the variable $z$ does not have a pole at $z=0$ anymore.\\[0.5em]
\indent\emph{Case $j\geq M+1,k\leq M$:} In this case, the sum over $x$ and then the residue at $w=v+1$ leads to
\begin{equation}
\langle \Phi^{n,t}_{n-j},\Psi^{n,t}_{n-k}\rangle =\frac{1}{2\pi\I}\oint_{\Gamma_0}\dx v (1+2v) (v(v+1))^{j-M-1}((v+1)(v+1-\alpha))^{M-k}=0
\end{equation}
because there is no pole at $v=0$ anymore.\\[0.5em]
\emph{Case $j\leq M,k\geq M+1$:} It is slightly more tricky to check the orthogonalization in this case. After the sum over $x$ and the residue at $w=z+1$, we get
\begin{equation}\label{eq3.19}
\begin{aligned}
&\langle \Phi^{n,t}_{n-j},\Psi^{n,t}_{n-k}\rangle\\
=& \frac{1}{(2\pi\I)^2}\oint_{\Gamma_{\alpha-1}}\dx v \frac{2v+2-\alpha}{((v+1)(v+1-\alpha))^{M-j+1}} \oint_{\Gamma_{0,v}}\dx z \frac{(z(z+1))^{M-k}(1+2z)}{(z-v)(z+v+1)}.
\end{aligned}
\end{equation}
This time, both poles at $z=0$ and $z=v$ contribute. One notices that the integrand in $z$ has four poles in the whole complex plane: $z=-1,0,-1-v,v$. By the change of variable $z=-1-w$, we have the identity
\begin{align}\label{eq3.18}
\oint_{\Gamma_{0,v}}\dx z \frac{(z(z+1))^{M-k}(1+2z)}{(z-v)(z+v+1)} &= \oint_{\Gamma_{-1,-1-v}}\dx w \frac{(w(w+1))^{M-k}(1+2w)}{(w-v)(w+v+1)} \nonumber \\
&=\frac12 \oint_{\Gamma_{-1,0,-1-v,v}}\dx z \frac{(z(z+1))^{M-k}(1+2z)}{(z-v)(z+v+1)}.
\end{align}
Since $M-k\leq -1$, the integrand is $\Or(1/z^3)$ at $z\to \infty$, thus the integral (\ref{eq3.18}) is zero, which implies then $\langle \Phi^{n,t}_{n-j},\Psi^{n,t}_{n-k}\rangle=0$.
\end{proofOF}

For our analysis we will not focus on the first $M-1$ particles, since it corresponds (up to a time-change) to the cases analyzed in previous works. Here we'll focus only on particles' positions of the ones with jump rate $1$. This is the reason why in what follows we write the kernel only for $n_1,n_2\geq M$. With the expressions of Lemma~\ref{LemmaPsiPhi} we can rewrite the kernel (\ref{eqKernelFinal}) in the following way.
\begin{prop}\label{PropStartingKernel}
For $n_1,n_2\geq M+1$, the kernel has the following expression
\begin{multline}\label{KernelIntRepr}
 K((n_1,t_1),x_1;(n_2,t_2),x_2) = - \hat\phi^{((n_1,t_1),(n_2,t_2))}(x_1,x_2)
 \\ +\widehat K^{(1)}((n_1,t_1),x_1;(n_2,t_2),x_2) +\widehat K^{(2)}((n_1,t_1),x_1;(n_2,t_2),x_2)
\end{multline}
where
\begin{equation}\label{phi_t}
 \widehat \phi^{((n_1,t_1),(n_2,t_2)}(x_1,x_2) = \frac{1}{2\pi i}\oint_{\Gamma_{0}}\frac{dw}{w}
 \frac{e^{(t_1-t_2)w} (w(w-1))^{n_1-n_2}}{w^{x_1+2n_1-x_2-2n_2}} \Id_{[(n_1,t_1)\prec (n_2,t_2)]},
\end{equation}
and
\begin{multline}\label{kernelK1}
 \widehat  K^{(1)}((n_1,t_1),x_1;(n_2,t_2),x_2)
 =\frac{1}{(2\pi\I)^2}\oint_{\Gamma_0}\dx v \oint_{\Gamma_{0,-v}}\frac{\dx w}{w}\frac{e^{t_1 w} (w(w-1))^{n_1-M}}{w^{x_1+2n_1-2M}} \\
  \times \frac{(1+v)^{x_2+2n_2-2M}}{e^{t_2(v+1)}(v(v+1))^{n_2-M}}\frac{(1+2v)}{(w+v)(w-v-1)}
\end{multline}
and
\begin{multline}\label{kernelK2}
\widehat   K^{(2)}((n_1,t_1),x_1;(n_2,t_2),x_2)
 =\frac{1}{(2\pi\I)^3}\oint_{\Gamma_{\alpha-1}}\dx v \oint_{\Gamma_{0,v}}\dx z \oint_{\Gamma_{0,\alpha-1-v}}\frac{\dx w}{w} \\
 \times \frac{e^{t_1w} (w(w-1))^{n_1-M} (w(w-\alpha))^M}{w^{x_1+2n_1-2M}}
  \frac{(1+z)^{x_2+2n_2-2M}}{e^{t_2(z+1)}(z(z+1))^{n_2-M}}  \\
  \times \frac{1}{((v+1)(v+1-\alpha))^M} \frac{(1+2z)(2v+2-\alpha)}{(z-v)(z+v+1)(w-1-v)(w+1-\alpha+v)}.
\end{multline}
\end{prop}

This proposition will be used in Section~\ref{SectTransition}.

\begin{proofOF}{Proposition~\ref{PropStartingKernel}}
For all $n_i\geq M$, we have $v_{n_i+1}=1$, thus (\ref{eqK0}) implies that $\phi^{((n_1,t_1),(n_2,t_2))}(x_1,x_2)$ in (\ref{eqK0}) equals $\widehat \phi^{((n_1,t_1),(n_2,t_2))}(x_1,x_2)$ plus the pole at $v=1$ (we return to it shortly). For the rest, we divide the sum over $k$ in (\ref{eqK0}) into the sum over $[1,\ldots,M]$ and the sum over $[M+1,\ldots,n_2]$. We define
\begin{equation}\label{eq3.26}
K^{(1)}((n_1,t_1),x_1;(n_2,t_2),x_2) = \sum_{k=M+1}^{n_2} \Psi^{n_1,t_1}_{n_1-k}(x_1) \Phi^{n_2,t_2}_{n_2-k}(x_2)
\end{equation}
and
\begin{equation}\label{eq3.27}
K^{(2)}((n_1,t_1),x_1;(n_2,t_2),x_2) = \sum_{k=1}^{M}\Psi^{n_1,t_1}_{n_1-k}(x_1) \Phi^{n_2,t_2}_{n_2-k}(x_2).
\end{equation}

Remark that $\Phi^{n_2,t_2}_{n_2-k}(x)=0$ for $k\geq n_2+1$. Therefore we can extend the sum in (\ref{eq3.26}) to infinity. Then, if we take $v$ small enough and $w$ large enough, satisfying $|v(v+1)|<|w(w-1)|$, we can take the sum inside the integrals. Explicitly, we get
\begin{multline}
K^{(1)}((n_1,t_1),x_1;(n_2,t_2),x_2) = \frac{1}{(2\pi\I)^2}\oint_{\Gamma_0} \dx v \oint_{\Gamma_{0,1}} \frac{\dx w}{w} \frac{(w(w-1))^{n_1-n_2}e^{t_1 w}}{w^{x_1+2n_1-2M}} \\ \times \frac{(1+v)^{x_2+2n_2-2M}(1+2v)}{e^{t_2(v+1)}}
\sum_{k=M+1}^\infty \frac{(w(w-1))^{n_2-k}}{(v(v+1))^{n_2-k+1}},
\end{multline}
where the integration contours have to satisfy $|v(v+1)|<|w(w-1)|$. Then, using
\begin{equation}
\sum_{k=M+1}^\infty \frac{(w(w-1))^{n_2-k}}{(v(v+1))^{n_2-k+1}} = \frac{(w(w-1))^{n_2-M}}{(v(v+1))^{n_2-M}} \frac{1}{(w-(1+v))(w+v)}
\end{equation}
we obtain (\ref{kernelK1}) plus the pole coming from $w=v+1$. This contribution cancels exactly with the contribution of the pole at $v=1$ of $\phi$.

We now show that $K^{(2)}=\widehat K^{(2)}$. For the computation of $K^{(2)}$, remark that the formula for $\Phi^{n_2,t_2}_{n_2-k}(x)$ used for $k\leq M$ gives exactly zero for $k\geq M+1$. The reason is that the pole at $v=\alpha-1$ disappears. Therefore we can use the integral representations for $k\leq M$ and extend the sum to $k=\infty$. Then, provided that $|(v+1)(v+1-\alpha)|\leq |w(w-\alpha)|$, we can exchange the sum and the integral, which gives
\begin{align}\label{eq3.30}
K^{(2)}((n_1,t_1),x_1;(n_2,t_2),x_2) &= \frac{1}{(2\pi\I)^3} \oint_{\Gamma_{0,\alpha}}\frac{dw}{w}\oint_{\Gamma_{\alpha-1}}\dx v \oint_{\Gamma_{0,v}}\dx z \frac{(w(w-1))^{n_1-M} e^{t_1 w}}{w^{x_1+2n_1-2M}} \notag \\
&\times \frac{(1+z)^{x_2+2n_2-2M}}{e^{t_2(z+1)}(z(z+1))^{n_2-M}} \frac{(1+2z)(2v+2-\alpha)}{(z-v)(z+v+1)} \notag \\
&\times \sum_{k=1}^\infty \frac{(w(w-\alpha))^{M-k}}{((v+1)(v+1-\alpha))^{M-k+1}}.
\end{align}
Then we substitute
\begin{equation}
\sum_{k=1}^\infty \frac{(w(w-\alpha))^{M-k}}{((v+1)(v+1-\alpha))^{M-k+1}} = \frac{(w(w-\alpha))^M}{((v+1)(v+1-\alpha))^M}\frac{1}{(w-1-v)(w+v+1-\alpha)}
\end{equation}
into (\ref{eq3.30}) to get (\ref{kernelK2}). The condition $|(v+1)(v+1-\alpha)|\leq |w(w-\alpha)|$ is satisfied for any $w\in \Gamma_{0,\alpha}$ if we choose $v$ sufficiently close to $\alpha-1$. Both poles $w=\alpha-1-v$ and $w=1+v$ lie inside $\Gamma_{0,\alpha}$. Thus, $K^{(2)}$ is given by $\widehat K^{(2)}$ but with the poles for $w=0,\alpha-1-v,1+v$. Consider the contribution coming from the pole at $w=v+1$, which is a simple residue. Computing this residue one immediately sees that the pole at $v=\alpha-1$ is not present anymore, thus the integral is zero.
\end{proofOF}

In the applications we'll use some special cases of the kernel too. In particular for $M=1$ and $M=\infty$ (with $n_i-M$ finite). Let us write the kernel explicitly in these cases.

\subsection{Special case: $M=1$}
\begin{cor}\label{CorMOne}
For $M=1$, the kernel has the following expression. For any $n_1,n_2\geq 1$,
\begin{multline}\label{eqKernelMOne}
K((n_1,t_1),x_1;(n_2,t_2),x_2) = - \widehat \phi^{((n_1,t_1),(n_2,t_2))}(x_1,x_2) \\
+\frac{1}{(2\pi\I)^2}\oint_{\Gamma_0}\dx v \oint_{\Gamma_{0,-v}}\frac{\dx w}{w}\frac{e^{t_1 w} (w-1)^{n_1-1}}{w^{x_1+n_1-1}}\frac{(1+v)^{x_2+n_2-1}}{e^{t_2(v+1)}v^{n_2-1}}\frac{(1+2v)}{(w+v)(w-v-1)}\\
+\frac{1}{2\pi\I}\oint_{\Gamma_0}\dx w \frac{e^{t_1 w} (w-1)^{n_1-1}}{w^{x_1+n_1}}\frac{1}{2\pi\I}\oint_{\Gamma_{0,\alpha-1}}\dx v \frac{(1+v)^{x_2+n_2-1}}{e^{t_2(v+1)} v^{n_2-1}}\frac{1+2v}{(v+1-\alpha)(v+\alpha)},
\end{multline}
with $\widehat \phi$ as in Proposition~\ref{PropStartingKernel}.
\end{cor}
\begin{proofOF}{Corollary~\ref{CorMOne}}
One simply substitutes for $M=1$ in the expression of Proposition~\ref{PropStartingKernel}. Then, the integral over $v$ around $\alpha-1$ is computed easily since it is a simple pole.
\end{proofOF}
The product structure of $K^{(2)}$ (the last term in (\ref{eqKernelMOne})) is straightforward if one looks back at its definition (\ref{eq3.27}). So, for $M=1$ we have a rank-one perturbation.

\subsection{Special case: $M=\infty$}
We want to get the $M\to\infty$ limit but with $n_i-M$ finite. Therefore, consider the kernel $K((M+n_1,t_1),x_1;(M+n_2,t_2),x_2)$ and take the $M\to\infty$ limit. We define
\begin{equation}
K_\infty((n_1,t_1),x_1;(n_2,t_2),x_2):=\lim_{M\to\infty}K((M+n_1,t_1),x_1;(M+n_2,t_2),x_2),
\end{equation}
and the limit kernel $K_\infty$ is given as follows.
\begin{cor}\label{CorMInfinity}
For $n_1,n_2\geq 1$, we have
\begin{multline}\label{eqCorMinfinity}
K_\infty((n_1,t_1),x_1;(n_2,t_2),x_2)=- \widehat \phi^{((n_1,t_1),(n_2,t_2))}(x_1,x_2) \\ +\frac{1}{(2\pi\I)^2}\oint_{\Gamma_0}\dx v \oint_{\Gamma_{0,-v}}\dx w \frac{e^{t_1 w} (w-1)^{n_1}}{w^{x_1+n_1+1}} \frac{(1+v)^{x_2+n_2}}{e^{t_2(v+1)}v^{n_2}}\frac{1+2v}{(w+v)(w-v-1)}\\
+\frac{-1}{(2\pi\I)^2} \oint_{\Gamma_{0}}\dx w \oint_{\Gamma_{0,\alpha-1-w}}\dx v  \frac{e^{t_1 w} (w-1)^{n_1}}{w^{x_1+n_1+1}} \frac{(1+v)^{x_2+n_2}}{e^{t_2(v+1)}v^{n_2}} \frac{1+2v}{(v+w+1-\alpha)(w-v-\alpha)}
\end{multline}
with $\widehat \phi$ as in Proposition~\ref{PropStartingKernel}.
\end{cor}
\begin{proofOF}{Corollary~\ref{CorMInfinity}}
The only not straightforward term is $K^{(2)}$. For $M\to\infty$ the pole at $w=0$ disappears and one just integrates out the simple pole at $w=\alpha-1-v$. The result does not depend on $M$ anymore, namely
\begin{equation}
\frac{-1}{(2\pi\I)^2} \oint_{\Gamma_{0}}\dx v \oint_{\Gamma_{0,v}}\dx z \frac{(1+z)^{x_2+n_2}}{e^{t_2(z+1)} z^{n_2}} \frac{e^{t_1(\alpha-1-v)} (\alpha-2-v)^{n_1}}{(\alpha-1-v)^{x_1+n_1+1}} \frac{1+2z}{(z-v)(z+v+1)}.
\end{equation}
Changing the variable $w=\alpha-1-v$ and then renaming $z$ with $v$ we get the result of the statement.
\end{proofOF}
Notice that for $\alpha=1$ the combination of the two integrals in (\ref{eqCorMinfinity}) is just the residue at $w=-v$, which is the kernel for alternating initial conditions already obtained in~\cite{BFPS06,BF07}. Moreover, for $n_1=n_2=n$, the kernel can also be seen as rank-$n$ perturbation of the kernel without the $K^{(1)}$ contribution. This kernel will be used explicitly in Section~\ref{SectWall}.

\subsection{Modified kernel useful for the shock region}
For the asymptotic analysis in the case of finite $M$ and $\alpha<1/2$, the shock situation, there is an interval around the shock for which the kernel has a diverging part in the $t\to\infty$ limit. This, however, does not mean that the system is ill-defined, because the distribution of the particles' positions is given by the Fredholm determinant of the kernel, not by the kernel itself. Indeed, we obtained a new kernel $K_{\rm shock}$ such that the Fredholm determinants agree. More importantly, in the new kernel (which is not a trivial conjugation of $K$) the divergence disappears. To define $K_{\rm shock}$ let us introduce the following function. Set
\begin{multline}
 Q_{n-j,w}^{n,t}(x)= \frac{1}{(2\pi \I)^2}\oint_{\Gamma_{\alpha-1}}\dx v \oint_{\Gamma_{w}}\dx z
 \frac{(1+z)^{x+2n-2M}}{e^{t(z+1)}(z(1+z))^{n-M}} \\
 \times \frac{1}{((v+1)(v+1-\alpha))^{M-j+1}} \frac{(1+2z)(2v+2-\alpha)}{(z-v)(z+v+1)},
\end{multline}
where $w$ is either $0,-1,v$. Then, for $j=1,\ldots,M$, we have
\begin{equation}
 \Phi_{n-j}^{n,t}(x) = Q_{n-j,0}^{n,t}(x) + Q_{n-j,v}^{n,t}(x).
\end{equation}
Define the new kernel $K_{\rm shock}$ for $n_1,n_2\geq M$ as follows:
\begin{multline}
K_{\rm shock}((n_1,t_1),x_1;(n_2,t_2),x_2)=- \hat\phi^{((n_1,t_1),(n_2,t_2))}(x_1,x_2) \\
+\widehat K^{(1)}((n_1,t_1),x_1;(n_2,t_2),x_2) +K_{\rm shock}^{(2)}((n_1,t_1),x_1;(n_2,t_2),x_2)
\end{multline}
where $K_{\rm shock}^{(2)}$ is defined as follows:
\begin{equation}
\begin{aligned}\label{eq4.38}
&K_{\rm shock}^{(2)}((n_1,t_1),x_1;(n_2,t_2),x_2) \\
&= \sum_{k=1}^M \Psi^{n_1,t_1}_{n_1-k}(x_1) Q^{n_2,t_2}_{n_2-k,v}(x_2)
+\sum_{k=1}^M \Psi^{n_1,t_1}_{n_1-k}(x_1) Q^{n_2,t_2}_{n_2-k,-1}(x_2)\Id_{[(n_1,t_1)\prec (n_2,t_2)]}.
\end{aligned}
\end{equation}
This means that for $(n_1,t_1)\prec (n_2,t_2)$ instead of the poles at $z=0,v$ in (\ref{eq3.14}) we have the poles at $z=-1,v$, and otherwise only the pole at $z=v$. The same changes in the poles will then occur in the triple integral representation (\ref{kernelK2}). This kernel will be useful for $\alpha<1/2$ because of the following Proposition.
\begin{prop}\label{PropModifiedKernel}
For given $a_1,\ldots,a_m\in\Z$, we have
\begin{equation}
 \det(\Id-\chi_a K\chi_a)_{\ell^2(\{(n_1,t_1),\ldots,(n_m,t_m)\}\times\Z)} =
 \det(\Id-\chi_a K_{\rm shock} \chi_a )_{\ell^2(\{(n_1,t_1),\ldots,(n_m,t_m)\}\times\Z)}.
\end{equation}
where $\chi_a((n_k,t_k),x)=\Id(x<a_k)$.
\end{prop}
Thus, by (\ref{joint}) the joint distribution of particles' positions can be computed with the kernel $K_{\rm shock}$ instead of $K$.

To prove Proposition~\ref{PropModifiedKernel} we'll use the following relations.
\begin{lem}\label{LemmaQrelations}
For $n_1,n_2\geq M$. Then
\begin{align}
 \sum_{x_1} Q_{n_1-j,-1}^{n_1,t_1}(x_1)K^{(1)}((n_1,t_1),x_1;(n_2,t_2),x_2)
 &= Q_{n_2-j,0}^{n_2,t_2}(x_2), \label{QK1} \\
 \sum_{x_1} Q_{n_1-j,-1}^{n_1,t_1}(x_1)K^{(2)}((n_1,t_1),x_1;(n_2,t_2),x_2)
 &= 0, \label{QK2} \\
 \sum_{x_1} Q_{n_1-j,-1}^{n_1,t_1}(x_1)
 \phi^{((n_1,t_1),(n_2,t_2))}(x_1,x_2)
 &= Q_{n_2-j,-1}^{n_2,t_2}(x_2). \label{Qphi}
\end{align}
\end{lem}
\begin{proofOF}{Lemma~\ref{LemmaQrelations}}
Recall the definition of $K^{(1)}$ and $K^{(2)}$, see (\ref{eq3.26})-(\ref{eq3.27}):
\begin{equation}
K^{(p)}((n_1,t_1),x_1;(n_2,t_2),x_2) = \sum_{k\in I_p} \Psi^{n_1,t_1}_{n_1-k}(x_1) \Phi^{n_2,t_2}_{n_2-k}(x_2)
\end{equation}
with $I_1=[M+1,\ldots,n_2]$ and $I_2=[1,\ldots,M]$. Then, we have to compute
\begin{equation}
\langle Q^{n_1,t_1}_{n_1-j,-1},\Psi^{n_1,t_1}_{n_1-k}\rangle\equiv \sum_{x\in\Z}Q^{n_1,t_1}_{n_1-j,-1}(x)\Psi^{n_1,t_1}_{n_1-k}(x),
\end{equation}
so some of the computations are very close to the ones we made for the orthogonalization in Lemma~\ref{LemmaPsiPhi}.

Let us start with (\ref{QK2}), i.e., $1\leq k \leq M$. Then, the expression of $Q$ is like $\Phi$ in (\ref{eq3.14}) but with $\Gamma_{0,v}$ replaced by $\Gamma_{-1}$. Thus, compare with (\ref{eq3.17}), we get
\begin{multline}
\langle Q^{n_1,t_1}_{n_1-j,-1},\Psi^{n_1,t_1}_{n_1-k}\rangle \\
 = \frac{1}{(2\pi\I)^2}\oint_{\Gamma_{\alpha-1}}\dx v \oint_{\Gamma_{-1}} \dx z \frac{((z+1)(z+1-\alpha))^{M-k}}{((v+1)(v+1-\alpha))^{M-j+1}} \frac{(1+2z)(2v+2-\alpha)}{(z-v)(z+v+1)}=0
\end{multline}
because the pole at $z=-1$ vanishes for $k\leq M$. This implies (\ref{QK2}).

Next we prove (\ref{QK1}), i.e., $k\geq M+1$. It is similar as before but with the $\Psi$ taken from (\ref{eq3.10}) instead of (\ref{eq3.14}). Then, compare with (\ref{eq3.19}), we have
\begin{multline}
\langle Q^{n_1,t_1}_{n_1-j,-1},\Psi^{n_1,t_1}_{n_1-k}\rangle \\= \frac{1}{(2\pi\I)^2}\oint_{\Gamma_{\alpha-1}}\dx v \oint_{\Gamma_{-1}}\dx z \frac{(z(z+1))^{M-k}(1+2z)}{(z-v)(z+v+1)}\frac{2v+2-\alpha}{((v+1)(v+1-\alpha))^{M-j+1}}.
\end{multline}
Thus, LHS of (\ref{QK1}) is given by
\begin{equation}\label{eq3.46}
 \sum_{k=M+1}^{n_2} \langle Q^{n_1,t_1}_{n_1-j,-1},\Psi^{n_1,t_1}_{n_1-k}\rangle \Phi_{n_2-k}^{n_2,t_2}(x_2).
\end{equation}
We use the fact that $\Phi^{n,t}_{n-k}(x)=0$ if $k>n$ to extend the sum to infinity. Then, provided $|w(w+1)|<|z(z+1)|$ we can take the sum inside the integrals; explicitly we get
\begin{multline}
(\ref{eq3.46})= \frac{1}{(2\pi\I)^3}\oint_{\Gamma_{\alpha-1}}\dx v \oint_{\Gamma_{-1}}\dx z \oint_{\Gamma_0}\dx w \frac{(1+2w)(1+w)^{x_2+n_2-M}}{e^{t_2(w+1)}w^{n_2-M}} \\ \times \frac{(1+2z)}{(z-v)(z+v+1)} \frac{2v+2-\alpha}{((v+1)(v+1-\alpha))^{M-j+1}}
\sum_{k\geq M+1} \frac{(w(w+1))^{k-M-1}}{(z(z+1))^{k-M}} \\
=\frac{1}{(2\pi\I)^2}\oint_{\Gamma_{\alpha-1}}\dx v \oint_{\Gamma_0}\dx w \frac{(1+w)^{x_2+n_2-M}}{e^{t_2(w+1)}w^{n_2-M}} \frac{1}{((v+1)(v+1-\alpha))^{M-j+1}}\\ \times \frac{(1+2w)(2v+2-\alpha)}{(w-v)(w+v+1)},
\end{multline}
where we integrated the pole at $z=w$ arising from the sum over $k$. This is however nothing else than $Q^{n_2,t_2}_{n_2-j,0}(x_2)$.

Finally, we need to verify (\ref{Qphi}). One divides the sum over $x_1$ in $[0,1,\ldots)$ and $(\ldots,-2,-1]$. Then use
\begin{equation}
\begin{aligned}
\sum_{x_1\geq 0} \left(\frac{1+z}{w}\right)^{x_1}=\frac{w}{w-1-z}&\textrm{ if } |1+z|<|w|, \\
\sum_{x_1< 0} \left(\frac{1+z}{w}\right)^{x_1}=-\frac{w}{w-1-z}&\textrm{ if } |1+z|>|w|.
\end{aligned}
\end{equation}
The two sums can be taken inside the integrals provided the contours satisfy once $|w|>|1+z|$ and the other time $|w|<|1+z|$. The integrands are the same up to a sign, which means that the net result of the sum is just the residue at $w=1+z$. Then (\ref{Qphi}) easily follows.
\end{proofOF}

With this result we can now proceed to the proof of Proposition~\ref{PropModifiedKernel}.
\begin{proofOF}{Proposition~\ref{PropModifiedKernel}}
In this proof we let $n$ stand for a pair $(n,t)$ for short
and write like $K_{n_1,n_2}(x_1,x_2)$ to represent
$((n_1,t_1),(n_2,t_2))$ block of $K$.
Let us define
\begin{equation}
 S_{n_1,n_2}(x_1,x_2) = \delta_{n_1,n_2}\delta_{t_1,t_2} \sum_{k=1}^M\Psi^{n_1,t_1}_{n_1-k}(x_1) Q^{n_1,t_1}_{n_1-k,-1}(x_2).
\end{equation}
Since $\Psi_{n-k}^{n,t}(x)=0$ for $x\leq 2M-n-k-1$ and $Q_{n-k,-1}^{n,t}(y)=0$ for $y\geq M-n$, it follows $\sum_{k=1}^M \Psi_{n-k}^{n,t}(x_1) Q_{n-k,-1}^{n,t}(x_2)=0$ if $x_1\leq M-n-1$ or $x_2\geq M-n$. Hence $S_{n,n}(x_1,x_2)=0$ for $x_2\geq x_1$, i.e., $S$ is lower triangular with diagonal being zero. Hence to prove the proposition, it is enough to show
\begin{equation}\label{eq3.50}
 (\Id-S)(\Id-K) = \Id-\tilde{K}.
\end{equation}
Since $S$ is block-diagonal, the $n_1,n_2$ block of (\ref{eq3.50}) writes
\begin{equation}\label{eq3.51}
(\Id-S)_{n_1,n_1}(\Id-K)_{n_1,n_2}=(\Id-K_{\rm shock})_{n_1,n_2},
\end{equation}
which is proven using the result of Lemma~\ref{LemmaQrelations} as follows. We use the notation $\overline K=K^{(1)}+K^{(2)}$ below.\\[0.5em]
\emph{Case $n_1=n_2$:} Then LHS of (\ref{eq3.51}) is
\begin{equation}\label{SKK11}
 (\Id-S_{n_1,n_1})(\Id-\overline{K}_{n_1,n_1}) = \Id-S_{n_1,n_1}-\overline{K}_{n_1,n_1}+S_{n_1,n_1}\overline{K}_{n_1,n_1}.
\end{equation}
We have $\overline{K}_{n_1,n_1}$ given by
\begin{equation}
 \overline{K}_{n_1,n_1} = K^{(1)}_{n_1,n_1} + \sum_{k=1}^M \Psi^{n_1,t_1}_{n_1-k} (Q^{n_1,t_1}_{n_1-k,0}+Q^{n_1,t_1}_{n_1-k,v}).
\end{equation}
and by Lemma~\ref{LemmaQrelations}
\begin{equation}
 S_{n_1,n_1} \overline{K}_{n_1,n_1} = \sum_{k=1}^M \Psi^{n_1,t_1}_{n_1-k} Q^{n_1,t_1}_{n_1-k,0}.
\end{equation}
Putting together these relations we obtain exactly RHS of (\ref{eq3.51}).\\[0.5em]
\emph{Case $n_1\neq n_2$ and $n_1\not\prec n_2$:} LHS of (\ref{eq3.51}) is in this case given by
\begin{equation}
-(\Id-S_{n_2,n_2})\overline K_{n_2,n_1}=-\overline K_{n_2,n_1}+S_{n_2,n_2}\overline K_{n_2,n_1}.
\end{equation}
Using
\begin{equation}
 \overline{K}_{n_2,n_1} = K_{n_2,n_1}^{(1)}+ \sum_{k=1}^M \Psi^{n_2,t_2}_{n_2-k} (Q^{n_1,t_1}_{n_1-k,0}+Q^{n_1,t_1}_{n_1-k,v}),
\end{equation}
and
\begin{equation}
 S_{n_2,n_2} \overline{K}_{n_2,n_1} = \sum_{k=1}^M \Psi^{n_2,t_2}_{n_2-k} Q^{n_1,t_1}_{n_1-k,0}
\end{equation}
we get
\begin{equation}
-(\Id-S_{n_2,n_2})\overline K_{n_2,n_1} = -K_{n_2,n_1}^{(1)}-\sum_{k=1}^M \Psi^{n_2,t_2}_{n_2-k} Q^{n_1,t_1}_{n_1-k,v}
\end{equation}
which is the claimed result.\\[0.5em]
\emph{Case $n_1\neq n_2$ and $n_1\prec n_2$:} in this case, LHS of (\ref{eq3.51}) is given by
\begin{equation}
 (\Id-S_{n_1,n_1})(-\overline{K}_{n_1,n_2}+\phi^{(n_1,n_2)}) = -\overline K_{n_1,n_2}+\phi^{(n_1,n_2)}+S_{n_1,n_1}\overline{K}_{n_1,n_2}-S_{n_1,n_1}\phi^{(n_1,n_2)}.
\end{equation}
This time Lemma~\ref{LemmaQrelations} tell us that
\begin{equation}
\begin{aligned}
 \overline{K}_{n_1,n_2} &= K_{n_1,n_2}^{(1)}+ \sum_{k=1}^M \Psi^{n_1,t_1}_{n_1-k} (Q^{n_2,t_1}_{n_2-k,0}+Q^{n_2,t_1}_{n_2-k,v}),  \\
 S_{n_1,n_1} \overline{K}_{n_1,n_2} &= \sum_{k=1}^M \Psi^{n_1,t_1}_{n_1-k} Q^{n_2,t_2}_{n_2-k,0}, \\
 S_{n_1,n_1} \phi^{(n_1,n_2)} &= \sum_{k=1}^M\Psi^{n_1,t_1}_{n_1-k} Q^{n_2,t_2}_{n_2-k,-1} .
\end{aligned}
\end{equation}
These relations imply the claimed result.
\end{proofOF}

\subsection{Special case: $M=1$}
For later use we explicitly state a corollary of Proposition~\ref{PropModifiedKernel}. For $M=1$, the extended kernel $K$ is given in Corollary~\ref{CorMOne}. Proposition~\ref{PropModifiedKernel} tell us that the Fredholm determinant can be also computed using the modified kernel $K_{\rm shock}$, which has the same expression as (\ref{eqKernelMOne}) but with the last term the integration for $v$ is around the poles $-1,\alpha-1$ instead of $0,\alpha-1$. In particular, for $(n_1,t_1)=(n_2,t_2)=(n,t)$, the kernel is given as follows.
\begin{cor}\label{CorMoneSlow}
For $M=1$, the one-point modified kernel is given by
\begin{equation}
K_{\rm shock}((n,t),x;(n,t),y)=K_{n,t}(x,y)+f(x)g(y),
\end{equation}
where
\begin{equation}
\begin{aligned}\label{eq4.1}
K_{n,t}(x,y)&=\frac{1}{(2\pi\I)^2}\oint_{\Gamma_0}\dx v \oint_{\Gamma_{0,-v}}\frac{\dx w}{w}\frac{e^{t w} (w-1)^{n-1}}{w^{x+n-1}}\frac{(1+v)^{y+n-1}}{e^{t(v+1)}v^{n-1}}\frac{(1+2v)}{(w+v)(w-v-1)} \\
f(x)&=\frac{1}{2\pi\I}\oint_{\Gamma_0}\dx w \frac{e^{t w} (w-1)^{n-1}}{w^{x+n}} \\
g(y)&=\frac{1}{2\pi\I}\oint_{\Gamma_{-1,\alpha-1}}\dx v \frac{(1+v)^{y+n-1}}{e^{t(v+1)} v^{n-1}}\frac{1+2v}{(v+1-\alpha)(v+\alpha)}.
\end{aligned}
\end{equation}
\end{cor}
This result, together with Proposition~\ref{PropModifiedKernel} will be employed in proving Proposition~\ref{PropShock}.

\section{Jam regime}\label{SectJam}
Consider the semi-infinite system with $1$ slow particle. By \emph{jam regime} we mean the following two situations in which particles with jump rate $1$ are slowed down by the slow particles:\\[0.5em]
(1) for $1/2\leq\alpha<1$: at large time $t$, the macroscopic density is continuous and has a plateau with density $1-\alpha$. The plateau corresponds to the first $(1-\alpha)^2 t$ particles moving with speed $\alpha$.\\[0.5em]
(2) for $0\leq \alpha<1/2$: in this case, the slow particle create a macroscopic shock and the density has a jump from $1/2$ to $1-\alpha$. Particles in the region of higher density move with speed $\alpha$. The shock has a drift velocity equal to $v_s=\alpha-1/2$ (i.e., it moves to the left).

With the results for case (2) we'll also be able to determine the law and the diffusion coefficient of the shock without introducing second-class particles.

\subsection{Fluctuations in the speed $\alpha$ region}
For large time $t$, particles with a particle number $n<\min\{\frac{1-\alpha}{2},(1-\alpha)^2\}t$ will move with the speed of the slow particles and will be very much correlated with the first $M$ slow particles. What happens is that the $M$th particle fluctuates according to the largest eigenvalue of ${\rm DBM}$. Intuitively, then the other particles have jump rate $1$, which is strictly larger than $\alpha$, so that they fill the gaps more rapidly than if the jump rate would have been $\alpha$. In doing so, their fluctuation will be well correlated with the last slow particle, just shifted in time. Therefore one might expect to see ${\rm DBM}$.

Before stating the result, we define the limit object we'll get in the large time limit. The matrix-valued (stationary) Ornstein-Uhlenbeck process on $M\times M$ hermitian matrices also known as Dyson's Brownian Motion, ${\rm DBM}$. It is the Markov process with transition density given by
\begin{equation}
 P(\tau_1,M_1;\tau_2,M_2) = \frac{1}{(2\pi (1-e^{-2(\tau_2-\tau_1)}))^{M}}
 \exp\left(-\frac{{\rm Tr}\left(M_2-e^{-(\tau_2-\tau_1)}M_1 \right)^2}{2(1-e^{-2(\tau_2-\tau_1)})}\right),
\end{equation}
for $\tau_2>\tau_1$, and the reference measure being flat over the independent entries of $M$, i.e., $dM=\prod_{i=1}^M dM_{i,i} \prod_{1\leq i<j\leq M} d\Re M_{i,j} d\Im M_{i,j}$.
Its finite-dimensional distributions are given by the following Fredholm determinant.

\begin{lem}\label{OU_kernel}
For any given $\tau_1<\ldots<\tau_m\in\R$, the joint distribution of the largest eigenvalue of the stationary ${\rm DBM}$ process are given by
\begin{equation}
 \Pb\left(\bigcap_{k=1}^m {\rm DBM}(\tau_k) \leq s_k \right) = \det(\Id-\chi_s K^{{\rm DBM}} \chi_s)_{L^2(\{\tau_1,\ldots,\tau_m\}\times \R)}
\end{equation}
where the kernel is given by
\begin{equation}
\begin{aligned}
\label{OU_K}
& K^{{\rm DBM}}(\tau_1,x_1;\tau_2,x_2) \\
&=-\frac{\exp\left(-{\displaystyle \frac{(x_2-x_1e^{-(\tau_2-\tau_1)})^2}{2(1-e^{-2(\tau_2-\tau_1)})}}\right)}{\sqrt{2\pi(1-e^{-2(\tau_2-\tau_1)})}}\Id_{[\tau_1<\tau_2]}
+  \sum_{k=1}^{M-1}e^{k(\tau_1-\tau_2)}p_k(x_1) p_k(x_2)e^{-x_2^2/2}
\end{aligned}
\end{equation}
where $p_k(x)=H_k(x/\sqrt{2}) \pi^{-1/4} 2^{-k/2} (k!)^{-1/2}$, and $H_k(x)$ is the standard Hermite polynomial of degree $k$ (see e.g.~\cite{KS96}).
\end{lem}
This result can be found in~\cite{Jo04} with a slightly different normalization, an extra $\sqrt{2}$ in the space variable.

\begin{prop}\label{PropJam} Let $\pi(\theta)$ be a real-valued function on $\R$ with $|\pi'|\leq 1$. Define the scaling
\begin{equation}
\begin{aligned}
t(\theta,T) &= (\pi(\theta)+\theta)\, T,\\
n(\theta,T) &= M+[(\pi(\theta)-\theta)T],
\end{aligned}
\end{equation}
where $T$ is the large parameter. For $0<n<\min\{\frac{1-\alpha}{2},(1-\alpha)^2\}t$, i.e.\ for $0<\pi(\theta)<\min\{\frac{2-\alpha}{1+\alpha},\frac{2-2\alpha+\alpha^2}{\alpha(2-\alpha)}\}\theta$, we are inside the region of speed $\alpha$. The rescaled process
\begin{equation}
 X_T(\theta) =\frac{x_{n(\theta,T)}(t(\theta,T))-(\alpha t -\frac{(n(\theta,T)-M)}{1-\alpha})}{-\sigma \sqrt{T}},
\end{equation}
where $\sigma^2\equiv \alpha (\pi(\theta)+\theta) - \frac{\alpha (\pi(\theta)-\theta)}{(1-\alpha)^2}$.
Then, in the large $T$ limit $X_T$ converges to the ${\rm DBM}$ process:
\begin{equation}\label{XOU}
\lim_{T\to \infty} X_T(\theta)={\rm DBM}(\tau(\theta)), \quad \textrm{with } \tau(\theta):=-\ln(\sigma)
\end{equation}
in the sense of finite dimensional distributions.
\end{prop}

Space-like paths include as particular cases: (a) fixed time with $t=T$ is obtained setting $\pi(\theta)=1-\theta$, and (b) fixed (tagged) particle with $n=T$ by setting $\pi(\theta)=1+\theta$. For more explanations about space-like paths see~\cite{BF07}.

\begin{remark}
In the following proof, as well as in the others on asymptotic analysis, we present only the most important ingredients. First of all we state explicitly the steep descent path used for the analysis and the local series expansions around the critical points (from where the non-vanishing term arises). These two are the building blocks for the convergence of the kernel on bounded sets, for more details on the procedure see e.g.\ Lemma 6.1 in~\cite{BF08}. We do not however prove convergence of the Fredholm determinants, for which bounds on moderate and large deviations are needed, for a simple example on how to proceed, see Lemma 6.2 in~\cite{BF08}.
\end{remark}

\begin{proofOF}{Proposition~\ref{PropJam}}
The result is obtained by analyzing the rescaled and conjugated kernel
\begin{equation}
K^{\rm resc}_T(\theta_1,\xi_1;\theta_2,\xi_2)=\frac{{\rm Conj}_2}{{\rm Conj}_1} \sigma_1 T^{1/2} K((n_1,t_1),x_1;(n_2,t_2),x_2)
\end{equation}
with $n_i:=n(\theta_i,T)$, $t_i:=t(\theta_i,T)$, and
\begin{equation}
x_i:=\alpha t_i-\frac{n_i}{1-\alpha}-\xi_i \sigma_i T^{1/2}.
\end{equation}
The conjugation factor is given by ${\rm Conj}_i:=e^{\alpha t_i} (\alpha-1)^{n_i}/\alpha^{x_i+n_i}$.
We use the kernel $K$ in Proposition~\ref{PropStartingKernel}.

\begin{figure}
\begin{center}
\psfrag{1}[c]{$1$}
\psfrag{0}[c]{$0$}
\psfrag{alpha}[c]{$\alpha$}
\psfrag{w2}[c]{$\omega_{+,2}$}
\psfrag{phi}[c]{Path for $\hat\phi$}
\psfrag{K1}[c]{Paths for $\widehat K^{(1)}$}
\psfrag{K2a}[c]{Paths for $\widehat K^{(2)}$, part (a)}
\psfrag{K2b}[c]{Paths for $\widehat K^{(2)}$, part (b)}
\psfrag{tildev}[cb]{$\tilde v$}
\psfrag{tildez}[cb]{$\tilde z$}
\psfrag{z}[cb]{$z$}
\psfrag{w}[cb]{$w$}
\includegraphics[height=7cm]{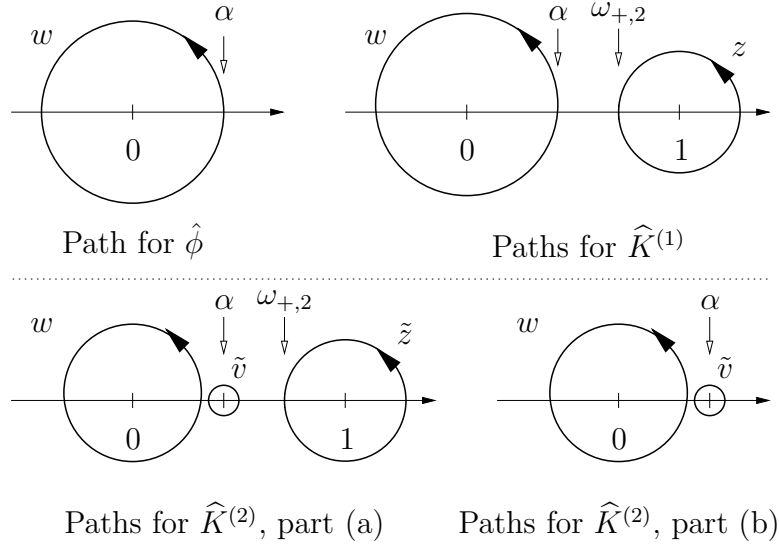}
\caption{Steep descents for the different terms in the kernel. The path $w$ passes either through $\alpha$ or at distance $\e T^{-1/2}$ from it.}
\label{FigSteepTrans1}
\end{center}
\end{figure}

Let us start with $\hat\phi$. In this proof, define the notation $a_i=\pi(\theta_i)-\theta_i$ and $u_i=\pi(\theta_i)+\theta_i$. Then, with the above scaling, for $(n_1,t_1)\prec (n_2,t_2)$,
\begin{equation}
\hat\phi^{((n_1,t_1),(n_2,t_2))}(x_1,x_2)=\frac{1}{2\pi\I}\oint_{\Gamma_0}\dx w e^{T g_0(w)+T^{1/2} g_1(w)+g_2(w)},
\end{equation}
where
\begin{equation}
\begin{aligned}
g_0(w)&= (u_1-u_2)(w-\alpha\ln(w))+(a_1-a_2)\left(\ln(w-1)+\frac{\alpha}{1-\alpha}\ln(w)\right),\\
g_1(w)&=(\xi_1\sigma_1-\xi_2\sigma_2)\ln(w),\\
g_2(w)&=-\ln(w).
\end{aligned}
\end{equation}
The condition $(n_1,t_1)\prec (n_2,t_2)$ means that $a_2-a_1\geq 0$ and $u_1-u_2\geq 0$ (at least one of the two inequalities being strict). The critical point for steep descent of $g_0(w)$ is at $w=\alpha$ and as steep descent path we use \mbox{$\Gamma_0=\{\alpha e^{\I y},y\in (-\pi,\pi]\}$}. Indeed,
\begin{equation}
\frac{\dx}{\dx y} \Re(g_0(w=\alpha e^{\I y}))=-\left((u_1-u_2)+\frac{a_2-a_1}{|w-1|^2}\right)\alpha \sin(y)
\end{equation}
which is negative for $\sin(y)>0$ and positive for $\sin(y)<0$, so the maximum of $\Re(g_0(w))$ is at $w=\alpha$ and the path $\Gamma_0$ is steep descent\footnote{For an integral $I=\int_\gamma \dx z e^{T f(z)}$, we
say that $\gamma$ is a steep descent path if (1) $\Re(f(z))$ is maximum at some $z_0\in\gamma$: $\Re(f(z))< \Re(f(z_0))$ for $z\in\gamma\setminus\{z_0\}$, and (2) $\Re(f(z))$ is monotone along $\gamma$ except at its maximum point $z_0$ and, if $\gamma$ is closed, at a point $z_1$ where the minimum of $\Re(f)$ is reached.}. By usual steep descent analysis, the relevant contribution of the integral comes from a \mbox{$\delta$-neighborhood} of $w=\alpha$. There, we can apply Taylor series:
\begin{equation}
\begin{aligned}
g_0(w)&=g_0(\alpha)-\frac{y^2}{2}(\sigma_1^2-\sigma_2^2)+\Or(y^3),\\
g_1(w)&=g_1(\alpha)+\I y(\xi_1\sigma_1-\xi_2\sigma_2)+\Or(y^2),\\
g_2(w)&=-\ln(\alpha)+\Or(y).
\end{aligned}
\end{equation}
Notice that $\exp(T g_0(\alpha)+T^{1/2} g_1(\alpha))={\rm Conj}_1/{\rm Conj}_2$. Then, deleting the corrections $\Or(\cdots)$ accounts for an error of order $\Or(T^{-1/2})$ times the leading term. The change of variable $y:=z T^{-1/2}$ implies that the leading term is given by
\begin{equation}
\begin{aligned}
&\frac{T^{-1/2}}{2\pi}\int_\R \dx z \exp\left(-(\sigma_1^2-\sigma_2^2) z^2/2+\I z (\xi_1\sigma_1-\xi_2\sigma_2)\right) \\
&=\frac{{\rm Conj}_1}{{\rm Conj}_2} \frac{1}{\sigma_1 T^{1/2}} \frac{1}{\sqrt{2\pi(1-\sigma_2^2/\sigma_1^2)}} \exp\left(-\frac{(\xi_1-\xi_2\sigma_2/\sigma_1)^2}{2(1-\sigma_2^2/\sigma_1^2)}\right).
\end{aligned}
\end{equation}

Now consider $\widehat K^{(1)}$. With the above scaling, and after the change of variable $v=z-1$, we get
\begin{equation}
\begin{aligned}
\label{eq5.16}
&\widehat K^{(1)}((n_1,t_1),x_1;(n_2,t_2),x_2) \\
&=\frac{1}{(2\pi\I)^2}\int_{\Gamma_1}\dx z\oint_{\Gamma_{0,1-z}}\dx w \frac{e^{T f_{0,1}(w)+T^{1/2} f_{1,1}(w)}}{e^{T f_{0,2}(z)+T^{1/2} f_{1,2}(z)}} \frac{2z-1}{(w+z-1)(w-z)w}
\end{aligned}
\end{equation}
with
\begin{equation}
\begin{aligned}
f_{0,i}(w)&=u_i (w-\alpha\ln(w)) + a_i \left(\ln(w-1)+\frac{\alpha}{1-\alpha}\ln(w)\right), \\
f_{1,i}(w)&=\xi_i\sigma_i \ln(w).
\end{aligned}
\end{equation}
Let us look for the critical points of $f_{0,i}$ and the steep descent paths. We have
\begin{equation}
\frac{\dx}{\dx w} f_{0,i}(w)=0 \iff \frac{(w-\alpha)(u_i(1-\alpha)(w-1)+a_i)}{w(w-1)(1-\alpha)}=0,
\end{equation}
that is, the critical points are
\begin{equation}
\omega_-=\alpha\quad\textrm{and}\quad \omega_{+,i}=1-\frac{a_i}{u_i (1-\alpha)}.
\end{equation}
Using the relation
\begin{equation}\label{eqBorder1}
0<a_i<u_i\min\{(1-\alpha)/2,(1-\alpha)^2\}
\end{equation}
one verifies the following relations:
\begin{equation}\label{eqBorder2}
0<\omega_-=\alpha<\omega_{+,i}<1,\quad \textrm{and}\quad \omega_{+,i}>1/2.
\end{equation}
As steep descent path we choose: $\Gamma_1=\{z=1-r e^{\I \psi}, r=1-\omega_{+,2}\}$ and $\Gamma_0=\{w=\alpha e^{\I \varphi}\}$. Let us check the steep descent property. We have
\begin{equation}
\frac{\dx}{\dx \varphi} \Re(f_{0,1}(w))=-\frac{\alpha u_1\sin(\varphi)} {|w-1|^2} \left(|w-1|^2-a_1/u_1\right)
\end{equation}
which is decreasing while moving away from the critical point $w=\alpha$, since by (\ref{eqBorder1}) the term in the parentheses is strictly positive. Also,
\begin{equation}
\frac{\dx}{\dx \psi} \Re(-f_{0,2}(z))=-\frac{r u_2\sin(\psi)}{|z|^2} \left(|z|^2-\alpha\left(1-\frac{a_2}{u_2(1-\alpha)}\right)\right)
\end{equation}
which is decreasing while moving away from the critical point $z=\omega_{+,2}$. Indeed, using (\ref{eqBorder1}) and (\ref{eqBorder2}) the term in the parentheses is strictly positive. Hence the integral (\ref{eq5.16}) is of order
\begin{equation}
\begin{aligned}
&\frac{{\rm Conj}_1}{{\rm Conj}_2} \frac{\exp\left(T\Re(f_{0,1}(\omega_-)-f_{0,1}(\alpha))\right)} {\exp\left(T\Re(f_{0,2}(\omega_{+,2})-f_{0,2}(\alpha))\right)}\\
=&\frac{{\rm Conj}_1}{{\rm Conj}_2} \exp\left(T\Re(f_{0,2}(\alpha)-f_{0,2}(\omega_{+,2}))\right)
= \frac{{\rm Conj}_1}{{\rm Conj}_2} \exp(-\delta T),\quad \delta>0.
\end{aligned}
\end{equation}
Indeed, $f_{0,2}(\omega_{+,2})>f_{0,2}(\alpha)$, because $\omega_{+,2}>\alpha$ and $\frac{\dx}{\dx z}f_{0,2}(z)>0$ on \mbox{$z\in (\alpha,\omega_{+,2})$}. Therefore in the $T\to\infty$ limit, the contribution of $K^{(1)}$ (conjugated and rescaled) goes to zero.

Finally, we need to consider $\widehat K^{(2)}$ of Proposition~\ref{PropStartingKernel}. We change the variables $v=\tilde v-1$ and $z=\tilde z-1$ in (\ref{kernelK2}) and get
\begin{multline}
 \widehat K^{(2)}((n_1,t_1),x_1;(n_2,t_2),x_2) = \frac{1}{(2\pi\I)^3}\oint_{\Gamma_\alpha}\dx \tilde v \oint_{\Gamma_{1,\tilde v}}\dx \tilde z\oint_{\Gamma_{0,\alpha-\tilde v}}\frac{\dx w}{w}\frac{(w(w-\alpha))^M}{(\tilde v(\tilde v-\alpha))^M}\\
 \times \frac{(2\tilde z-1)(2\tilde v-\alpha)}{(\tilde z+\tilde v-1)(w-\tilde v)(\tilde z-\tilde v)(w+\tilde v-\alpha)} \frac{e^{T f_{0,1}(w)+T^{1/2}f_{1,1}(w)}}{e^{T f_{0,2}(\tilde z)+T^{1/2} f_{1,2}(\tilde z)}}.
\end{multline}
We have the following two contributions:\\
\textit{(a): $\tilde z$ around $1$.} From the above analysis on the steep descent paths, we choose $|\tilde v-\alpha|=\e/2$ for $\e$ small enough, $\{w=(\alpha-\e) e^{\I \varphi}\}$ and $\{z=1-r e^{\I \psi}, r=1-\omega_{+,2}\}$. Since $f_{0,2}(\omega_{+,2})>f_{0,2}(\alpha)$, we can choose $\e$ small enough such that the overall contribution is $\frac{{\rm Conj}_1}{{\rm Conj}_2} \exp(-\delta' T)$ for some $\delta'>0$. Therefore in the $T\to\infty$ limit, this contribution of $\widehat K^{(2)}$ (conjugated and rescaled) goes to zero. \\
\textit{(b): $\tilde z$ around $\tilde v$.} Integrating out the simple pole $\tilde z=\tilde v$ we get a contribution equal to
\begin{equation}\label{eq5.26}
\frac{1}{(2\pi\I)^2}\oint_{\Gamma_\alpha}\dx \tilde v \oint_{\Gamma_{0,\alpha-\tilde v}}\frac{\dx w}{w} \frac{e^{T f_{0,1}(w)+T^{1/2}f_{1,1}(w)}}{e^{T f_{0,2}(\tilde v)+T^{1/2} f_{1,2}(\tilde v)}} \frac{(w(w-\alpha))^M}{(\tilde v(\tilde v-\alpha))^M}\frac{2\tilde v-\alpha}{(w-\tilde v)(w+\tilde v-\alpha)}.
\end{equation}
The integration paths are now chosen as $\Gamma_{\alpha}=\{|\tilde v-\alpha|=R T^{-1/2}\}$ and $\Gamma_{0,\alpha-\tilde v}=\{w=(\alpha-LT^{-1/2}) e^{\I y}, L>R\}$. With the above computations we have that $\Gamma_{0,\alpha-\tilde v}$ is a steep descent path, so that its leading contribution comes from a $T^{-1/2}$ neighborhood of $y=0$. There we can use Taylor series:
\begin{equation}
\begin{aligned}
f_{0,i}(w)&=f_{0,i}(\alpha)+\frac{(w-\alpha)^2}{2\alpha^2}\sigma_i^2+\Or((w-\alpha)^3),\\
f_{1,i}(w)&=f_{1,i}(\alpha)+\frac{\xi_i\sigma_i(w-\alpha)}{\alpha} + \Or((w-\alpha)^2),\\
(w(w-\alpha))^{M} &= \alpha^M(w-\alpha-1)^{M}(1+\Or((w-\alpha))).
\end{aligned}
\end{equation}
Denoting $\tilde v=\alpha(1+V\sigma_1^{-1} T^{-1/2})$ and $w=\alpha(1+ W\sigma_1^{-1} T^{-1/2})$ we have that the leading term of (\ref{eq5.26}) for large $T$ is given by
\begin{equation}
\frac{{\rm Conj}_1}{{\rm Conj}_2} \frac{1}{\sigma_1 T^{1/2}} \frac{1}{(2\pi\I)^2}\oint_{|V|=R}\dx V \int_{-L+\I\R}\dx W \frac{W^M}{V^M}\frac{1}{W-V} \frac{e^{W^2/2+W\xi_1}}{e^{V^2(\sigma_2/\sigma_1)^2/2+V\xi_2\sigma_2/\sigma_1}}
\end{equation}
with $L>R$. By the change of variable $W\to -W$ and $V\to -V$ we obtain (up to factors $\sqrt{2}$ due to the different space-scaling) the extended Hermite kernel, see (2.13) of~\cite{Jo04}, which can then be rewritten in terms of Hermite polynomials.
\end{proofOF}

\subsection{Fluctuations around the shock}\label{subsectShock}
In this section we consider $\alpha<1/2$ and focus around the shock position. We consider the case of $M=1$ slow particle, since it is also physically the more natural situation. Indeed, the system with one slow particle is equivalent to having stationary initial condition on $\Z_+$. The shock position, which at time $t$ will be around position $v_s t=(\alpha-1/2)t$, fluctuates on the $t^{1/2}$-scale. How does it happen? The fluctuations of particles before entering the shock region live on a $t^{1/3}$~scale, thus viewed from the $t^{1/2}$-scale, these particles essentially do not fluctuate. So, one has some probability that the particle is not inside the shock region, which will then be very well localized, and if the particle is in the shock region, then it follows fluctuations of the slow particles. The occurrence of these two distinct intermediate scales allows us, in particular, to determine the diffusion coefficient of the shock. The result agrees with the argument in~\cite{SpohnBook91} modified appropriately for our situation, see below.

The first result was stated in Proposition~\ref{PropShock} and in order to prove it we recall the following result from~\cite{BFS07}.
\begin{lem}\label{lemmaShock}
Consider the kernel without the slow particle, i.e., $K_{n,t}$ defined in (\ref{eq4.1}), and the rescaling
\begin{equation}\label{eqLemmSchock}
n=\nu t,\quad x_i=\tfrac12 t-2n-\zeta_i t^{1/3},\quad\textrm{with }\nu>1/4.
\end{equation}
Then, uniformly for $\zeta_i$ in a bounded set,
\begin{equation}
\lim_{t\to\infty} t^{1/3} K_{n,t}(x_1,x_2)\equiv
K_{\Af}(\zeta_1,\zeta_2),
\end{equation}
where $K_{\Af}$ is the Airy$_1$ kernel~\cite{Sas05,BFPS06,Fer07}.
\end{lem}
With $\equiv$ we mean equivalent, since indeed to get a well-defined limit one has to do a conjugation of the kernel $K_{n,t}$.

It is also known~\cite{FS05b} that
\begin{equation}
F_1(2s)=\det(\Id-K_{\Af})_{L^2((s,\infty),\dx x)},
\end{equation}
with $F_1$ the GOE Tracy-Widom distribution function~\cite{TW96}.

\begin{proofOF}{Proposition~\ref{PropShock}} Let us start with $\xi>0$. From Proposition~\ref{PropModifiedKernel} and Corollary~\ref{CorMoneSlow} we have
\begin{equation}
\begin{aligned}
\Pb(x_n(t)\geq x)&=\det(\Id-\chi_x K_{n,t}\chi_x -\chi_x f\cdot g\chi_x )\\
&=\det(\Id-\chi_x K_{n,t}\chi_x)(1-(g\chi_x ,(\Id-\chi_x K_{n,t}\chi_x)^{-1} \chi_x f))
\end{aligned}
\end{equation}
with $K_{n,t}$, $f$ and $g$ being defined in (\ref{eq4.1}), and $\chi_x=\Id_{(-\infty,x)}$. Compare the scaling (\ref{eqShock}) and (\ref{eqLemmSchock}): $\nu\leftrightarrow \frac{1-\alpha}{2}+\eta t^{-1/2}$ and $\zeta_i\leftrightarrow\xi t^{1/6}$. So, for any $\xi>0$ we will actually focus on the upper tail of the Airy$_1$ kernel and of the related distribution, i.e., $K_{n,t}(x,y)\to 0$ as $t\to\infty$ (with conjugation), and
\begin{equation}
\lim_{t\to\infty}\det(\Id-\chi_x K_{n,t}\chi_x)=1.
\end{equation}
For the second term, we use
\begin{equation}
(\Id-\chi_x K_{n,t}\chi_x)^{-1} =\Id+\chi_x K_{n,t} \chi_x(\Id-\chi_x K_{n,t}\chi_x)^{-1}.
\end{equation}
Thus
\begin{equation}
(g\chi_x ,(\Id-\chi_x K_{n,t}\chi_x)^{-1} \chi_x f)=(g\chi_x,f)+(g\chi_x,(\Id-\chi_x K_{n,t}\chi_x)^{-1} \chi_x K_{n,t}\chi_x f)
\end{equation}
and the fact that $K_{n,t}\to 0$ implies then that the last term goes to zero. So
\begin{equation}
\lim_{t\to\infty}\Pb(x_n(t)\geq x) = \lim_{t\to\infty} 1-(g\chi_x,f) =\lim_{t\to\infty} (g\tilde \chi_x,f),\quad \tilde \chi_x=\Id_{[x,\infty)}.
\end{equation}
In the last step we used the orthogonality between $g$ and $f$, namely $(g,f)=1$. Under the scaling (\ref{eqShock}), $x(\xi)+n\sim \tfrac12\alpha t>0$, which means that the pole at $v=-1$ in the function $g(y)$ defined in (\ref{eq4.1}) vanishes. Therefore,
\begin{equation}
g(x)=\frac{\alpha^{x+n-1}}{e^{\alpha t} (\alpha-1)^{n-1}}
\end{equation}
and then
\begin{equation}
\begin{aligned}
(g\tilde \chi_x,f) &= \frac{1}{2\pi\I}\oint_{|w|>\alpha}\dx w e^{t(w-\alpha)}\left(\frac{(w-1)\alpha}{(\alpha-1)w}\right)^{n-1}\sum_{y\geq x(\xi)}\alpha^y/w^{y+1}  \\
&=\frac{1}{2\pi\I}\oint_{|w|>\alpha}\dx w e^{t(w-\alpha)}\left(\frac{(w-1)\alpha}{(\alpha-1)w}\right)^{n-1}\left(\frac{\alpha}{w}\right)^{x(\xi)}\frac{1}{w-\alpha}  \\
&\equiv \frac{1}{2\pi\I}\oint_{|w|>\alpha}\dx w  e^{F(w)} \frac{1}{w-\alpha}.
\end{aligned}
\end{equation}
The steep descent path of $F$ passes by the saddle point at $w=\alpha$, but since it is a pole we have just to deform locally on a $t^{-1/2}$~scale to pass on its right. The leading contribution is coming from a $t^{-1/2}$-neighborhood of the $w=\alpha$. Setting $w=\alpha+\I y \alpha t^{-1/2}$, we get
\begin{equation}
F(w)=-\sigma^2 y^2/2+\I y(\xi+\xi_c)+\Or(y^3 t^{-1/2}).
\end{equation}
The $\Or(y^3 t^{-1/2})$ term is controlled by the quadratic term, and in the end we obtain
\begin{equation}
\begin{aligned}
\lim_{t\to\infty}(g\tilde \chi_x,f)&=\frac{1}{2\pi\I}\int_{\R+\I\e} \dx y \frac{e^{-\sigma^2 y^2/2+\I y(\xi+\xi_c)}}{y},\quad \forall \e>0 \\
&=\frac{1}{\sqrt{2\pi\sigma^2}}\int_{-\infty}^{\xi+\xi_c}\dx z \exp\left(-z^2/(2\sigma^2)\right).
\end{aligned}
\end{equation}

Now consider $\xi<0$. We first use a probabilistic argument. It is quite clear (by a simple coupling argument) that
\begin{equation}
\Pb(x_n(t)\geq x)\leq \widetilde \Pb(x_n(t)\geq x)
\end{equation}
where $\widetilde \Pb$ is the measure of the system without the slow particle.
For the system without slow particle we have the result of Lemma~\ref{lemmaShock}, which tells us that
\begin{equation}\label{eq5.30}
\widetilde \Pb(x_n(t)\geq x) =\det(\Id-\chi_x K_{n,t} \chi_x)\to 0,\quad t\to\infty.
\end{equation}
The reason is that $\xi<0$ corresponds to $\zeta_i=\xi t^{1/6}\to -\infty$ as $t\to\infty$. Then (\ref{eq5.30}) follows from the non-degeneracy of the distribution $F_1$ (no mass is lost at $-\infty$).
\end{proofOF}

It is a bit more natural to look at the fluctuations with respect to the dashed line in Figure~\ref{FigureShock}. Then, the result of Proposition~\ref{PropShock} rewrites as follows. Let
\begin{equation}
F(\xi):=\lim_{t\to\infty}\Pb(x_{n=[(1-\alpha)t/2+\eta t^{1/2}]}(t)\geq \alpha t-n-\xi t^{1/2}).
\end{equation}
Then, $F(\xi)$ has a jump at $\xi=\xi_c$, namely
\begin{equation}\label{eqShockDensity}
F'(\xi)=\frac{1}{\sqrt{2\pi\sigma^2}}\exp(-\xi^2/2\sigma^2) \Id_{[\xi>\xi_c]}+\delta(\xi-\xi_c)\frac{1}{\sqrt{2\pi\sigma^2}}\int_{\xi_c}^\infty \dx y \exp(-y^2/2\sigma^2).
\end{equation}
This can be used to determine the diffusion coefficient of the shock without having to identify it with second class particles. When $\alpha<1/2$, the macroscopic density has a jump from $1/2$ to $1-\alpha$. As we saw in Proposition~\ref{PropJam}, before the shock the fluctuations becomes asymptotically $F_1$-distributed on a $t^{1/3}$-scale, while inside the shock region are Gaussian on the $t^{1/2}$-scale. Thus the position of the shock itself is localized on the $t^{1/2}$-scale, see Figure~\ref{FigureShock} for an illustration.

The question we want to address is how to determine its law and in particular its diffusion coefficient. Denote by $x_{\rm shock}(t)$ the position of the shock at time~$t$.
\begin{prop}\label{PropShockDiffusionCoeff}
In the large time limit, the shock is Gaussian distributed with diffusion coefficient $D$ given by
\begin{equation}\label{eq5.47}
D=\frac{\alpha(1-\alpha)}{1/2-\alpha}.
\end{equation}
In other words,
\begin{equation}
\lim_{t\to\infty}\Pb(x_{\rm shock}(t)\geq (\alpha-1/2)t-\nu t^{1/2}) = \frac{1}{\sqrt{2\pi D}}\int_{\nu}^\infty \dx x \exp(-x^2/2D).
\end{equation}
\end{prop}
\begin{proofOF}{Proposition~\ref{PropShockDiffusionCoeff}}
To prove the result we first have to understand what Proposition~\ref{PropShock} says. Consider the particle with number $n$ and look at position $x$ rescaled as in (\ref{eqShock}). The condition $\xi>\xi_c$ means that $x$ is on the left of the dotted line of Figure~\ref{FigureShock} by $(\xi-\xi_c)t^{1/2}$. Moreover, before reaching the shock, particles fluctuate only on a $t^{1/3}$-scale away from the dotted line. Thus, for $\xi>\xi_c$, $x_n(t)<x$ implies that particle $n$ already reached the shock. On the other hand, if particle $n$ did not reach the shock region yet, then (on the $t^{1/2}$~scale) it has to be on the dotted line (can not be farther to the right because of (\ref{eqShock3})). Therefore, the probability that particle $n$ has not yet reached the shock (i.e.\ $x_{\rm shock}(t)>x_n(t)$) is equal to the mass at $\xi=\xi_c$. Thus, from (\ref{eqShockDensity}) it follows that
\begin{align}\label{eq5.49}
&\lim_{t\to\infty}\Pb(x_{\rm shock}(t)\geq (\alpha-1/2)t-\nu t^{1/2}) \\
&= \frac{1}{\sqrt{2\pi\sigma^2}}\int_{(1/2-\alpha)\nu/(1-\alpha)}^\infty \dx y \exp(-y^2/2\sigma^2) = \frac{1}{\sqrt{2\pi D}}\int_{\nu}^\infty \dx x \exp(-x^2/2D) \notag
\end{align}
after a change of variable.
\end{proofOF}

\begin{remark}
The above argument is quite flexible and one could extend to the case of $M$ slow particles instead of only one. We expect the following. Proposition~\ref{PropShock} would be similar up to the distribution in (\ref{eqShock2}) changed from Gaussian into the ${\rm GUE}(M)$ (the distribution of the largest eigenvalue of $M\times M$ GUE matrices) and the shock will have a ${\rm GUE}(M)$-distribution with appropriate parameter, by the change of variable as in (\ref{eq5.49}).
\end{remark}

This result can also be explained with an heuristic argument, following arguments in~\cite{SpohnBook91}. In the continuum limit the particle
density $\rho_t(x)$ is described by the viscous Burgers equation with noise (see (5.37) of~\cite{SpohnBook91}),
\begin{equation}
 \frac{\partial}{\partial t} \rho_t(x)
 +
 \frac{\partial}{\partial x}(\rho_t(x)(1-\rho_t(x)))
 =
 \epsilon \nu \frac{\partial^2}{\partial x^2} \rho_t(x)
 -
 \sqrt{\epsilon \nu} \frac{\partial}{\partial x} J_t(x).
\label{Burgers}
\end{equation}
Here $\epsilon$ is the lattice constant, $\nu$ is the diffusion
constant and $J_t(x)$ is the random current.
The initial condition is divided into two parts,
\begin{equation}
 \rho_0(x) = \rho_s(x) + \sqrt{\epsilon} \xi(x).
\end{equation}
Here $\rho_s$ is the deterministic part,
\begin{equation}
 \rho_s(x) = \left\{\begin{array}{ll}
          \rho_-:=1/2, & x<0, \\
              \rho_+:=1-\alpha, & x>0,
         \end{array}\right.
\end{equation}
and $\xi(x)$ takes into account the randomness in the initial
conditions for $x>0$,
\begin{equation}
 \langle \xi(x) \xi(x') \rangle =
\left\{\begin{array}{ll}
  0,  & x<0, \\
  \rho_+(1-\rho_+) \delta(x-x'), & x>0.
 \end{array}\right.
\label{cov}
\end{equation}
Note that in our present case there is no randomness for $x<0$.
The solution to (\ref{Burgers}) is of the form,
\begin{equation}
 \rho_t(x) = \rho_s( x-v_s t-\sqrt{\epsilon D}b(t)) + O(\sqrt{\epsilon}),
\end{equation}
with $b(t)$ is the standard Brownian Motion. The shock front remains sharp but its center performs the Brownian Motion.
The diffusion coefficient of the shock $D$ is of our interest.

Let us suppose that the initial density fluctuations move with
constant velocity towards the shock and that this is the source
of randomness of the shock location. At time $t$ the particle
density fluctuations starting from the region $[v_s t,-v_s t]$
have arrived at the shock so that $\int_{v_s t}^{-v_s t}\xi(x)dx$
represents the excess amount of particles comparing to the
deterministic part. Since the difference of the density to the
left and the right is $\rho_+-\rho_-$, we would have
\begin{equation}
 \sqrt{D}b(t) = \frac{1}{\rho_+-\rho_-} \int_{v_s t}^{-v_s t} \xi(x) dx.
\end{equation}
Using (\ref{cov}) we get
\begin{equation}
 D = \frac{\rho_+(1-\rho_+)}{\rho_+-\rho_-} = \frac{\alpha(1-\alpha)}{1/2-\alpha}.
\end{equation}
This is the same as (\ref{eq5.47}).

\section{Transition processes}\label{SectTransition}
In this section, we first focus around the critical parameter $\alpha=1/2$ and later on the Airy$_2$ to $\rm DBM(M)$ transition. For $\alpha=1/2$, on a macroscopic scale the density is constant and equal to $1/2$. However, the fluctuations to the left of the origin live on the $t^{1/3}$~scale, while on the right they live on the $t^{1/2}$~scale. Here we consider $\alpha-1/2=\Or(t^{-1/3})$ and $n-t/4=\Or(t^{2/3})$. We keep $M$ fixed and finite.

As before, we are not obliged to stay on a fixed time, but we can consider a space-like path described by a function $\pi(\theta)$ with $|\pi'|\leq 1$. Consider the space-like setting as in Proposition~\ref{PropJam},
\begin{equation}\label{eq6.1}
\begin{aligned}
t(\tau,T) &= (\pi(\theta-\tau T^{-1/3})+\theta-\tau T^{-1/3})\, T,\\
n(\tau,T) &= M+[\pi(\theta-\tau T^{-1/3})-(\theta-\tau T^{-1/3})]\, T,\\
\end{aligned}
\end{equation}
with $\theta>0$ \emph{fixed} and\footnote{This does not mean that the function $\widetilde \theta\mapsto \pi(\widetilde \theta)$ is identically equal to $5\widetilde \theta/3$, only that at $\widetilde \theta=\theta$ its value is equal to $5\theta/3$.} $\pi(\theta)=5\theta/3$. This ensures that macroscopically we focus at the transition region, which for $\alpha=1/2$ is around $n=t/4$.

Here we consider $\alpha$ not necessarily exactly equal to $1/2$. Instead, let us define
\begin{equation}\label{eq5.2}
\alpha=\tfrac12(1+\kappa T^{-1/3}).
\end{equation}
Then, the rescaled process of particle position is given by
\begin{equation}\label{eq6.3}
X_T(\tau) =
\left\{\begin{array}{ll}
{\displaystyle \frac{x_{n}(t)-(\tfrac12 t-2(n-M))}{-T^{1/3}}}, &\textrm{if }n\geq t/4,\\[1em]
{\displaystyle \frac{x_n(t)-(t-2\sqrt{t(n-M)})}{-T^{1/3}}},&\textrm{if }n\leq t/4.
\end{array}\right.
\end{equation}
In the large-$T$ limit, $X_T$ will converge to a well-defined limit process, ${\cal A}_{2\to 1,M,\kappa}$, which we now define.

\begin{figure}[t]
\begin{center}
\psfrag{g1t}[c]{$\tilde\gamma_1$}
\psfrag{g2t}[c]{$\tilde\gamma_2$}
\psfrag{mg2t}[r]{$-\tilde\gamma_2$}
\psfrag{g1}[c]{$\gamma_1$}
\psfrag{g2}[c]{$\gamma_2$}
\psfrag{gk}[c]{$\Gamma_\kappa$}
\psfrag{k}[c]{$\kappa$}
\includegraphics[height=4cm]{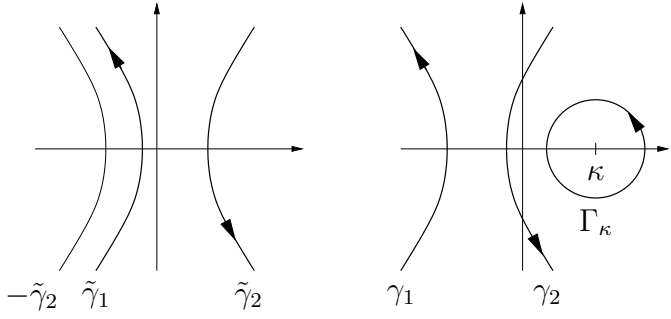}
\caption{Illustration of the integration paths defining the kernel $K^{\rm trans}_{M,\kappa}$.}
\label{FigurePathTransition}
\end{center}
\end{figure}
\begin{defin}\label{DefTransition}
Let us set
\begin{equation}
\tilde s_i=\left\{\begin{array}{ll}
s_i,&\textrm{if }\tau_i\geq 0,\\
s_i-\tau_i^2,&\textrm{if }\tau_i\leq 0.
\end{array}\right.
\end{equation}
The process ${\cal A}_{2\to 1,M,\kappa}$ is the process with $m$-point distributions at $\tau_1<\tau_2<\ldots <\tau_m$ given by the Fredholm determinant,
\begin{equation}
 \Pb\left( \bigcap_{k=1}^m\left\{{\cal A}_{2\to 1,M,\kappa}(\tau_k)\leq s_k\right\} \right)
 = \det \left(\Id-\chi_s K_{M,\kappa}^{\rm trans}\chi_s \right)_{L^2(\{\tau_1,\ldots,\tau_m\}\times \mathbb{R})}
\end{equation}
where $\chi_s(\tau_k,x)=\Id(x>s_k)$. The kernel is defined by
\begin{equation}
 K^{{\rm trans}}_{M,\kappa}(\tau_1,s_1;\tau_2,s_2) = K^{{\rm trans}(1)}(\tau_1,s_1;\tau_2,s_2)
 + K_{M,\kappa}^{{\rm trans}(2)}(\tau_1,s_1;\tau_2,s_2),
\end{equation}
with
\begin{multline}\label{Ktr0}
 K^{{\rm trans}(1)}(\tau_1,s_1;\tau_2,s_2) = -\frac{1}{\sqrt{4\pi (\tau_2-\tau_1)}}
 \exp\left(-\frac{(\tilde s_2-\tilde s_1)^2}{4(\tau_2-\tau_1)}\right) \Id_{[\tau_2>\tau_1]} \\
 \quad + \frac{1}{(2\pi\I)^2} \int_{\tilde\gamma_2}\dx w_2 \int_{\tilde \gamma_1}\dx w_1
 \frac{e^{w_2^3/3+\tau_2 w_2^2-\tilde s_2w_2}}{e^{w_1^3/3+\tau_1 w_1^2-\tilde s_1w_1}}
 \frac{2w_2}{(w_1-w_2)(w_1+w_2)},
\end{multline}
and
\begin{multline}\label{Ktr1}
 K_{M,\kappa}^{{\rm trans}(2)}(\tau_1,s_1;\tau_2,s_2) = \frac{1}{(2\pi\I)^3}
 \oint_{\Gamma_{\kappa}}\dx u \int_{\gamma_2}\dx w_2 \int_{\gamma_1}\dx w_1
 \frac{e^{w_2^3/3+\tau_2 w_2^2-\tilde s_2w_2}}{e^{w_1^3/3+\tau_1 w_1^2-\tilde s_1w_1}} \\
 \quad\times \frac{2w_2}{(w_2-u)(w_2+u)(w_1-u)}\left(\frac{w_1-\kappa}{u-\kappa}\right)^M.
\end{multline}
Here $\tilde\gamma_2,\gamma_2: e^{\pi\I/3} \infty\rightarrow e^{-\pi\I/3}\infty$, $\tilde\gamma_1,\gamma_1: e^{-2\pi\I/3} \infty\rightarrow e^{2\pi\I/3}\infty$, and $\Gamma_{\kappa}$ goes around only the pole at $u=\kappa$ anticlockwise. Moreover, $-\tilde \gamma_2\subset \tilde \gamma_1$, and $\gamma_1,\gamma_2$ passes on the left of $\Gamma_\kappa$ (see Figure~\ref{FigurePathTransition} for an illustration).
\end{defin}

\begin{thm}\label{ThmTransition}
The process $X_T$ defined in (\ref{eq6.3}) converges to the process ${\cal A}_{2\to 1,M,\kappa}$, more precisely
\begin{equation}
\lim_{T\to\infty} X_T(\tau) = S_v {\cal A}_{2\to 1,M,\kappa S_v}(\tau/S_h)
\end{equation}
in the sense of finite-dimensional distributions. The scaling coefficients $S_h$ and $S_v$ are given by
\begin{equation}
S_v=\left(\frac{4\theta}{3}\right)^{1/3},\quad S_h=\frac{4}{5-3\pi'(\theta)} S_v^2.
\end{equation}
\end{thm}
In the fixed time case, $t=T$, we have $\pi(\theta)=1-\theta$ and $\pi(\theta)=5\theta/3$, from which $\theta=3/8$, i.e., $S_v=2^{-1/3}$ and $S_h=2^{-5/3}$.

\begin{remark}
When $M=0$ we have $K_{M,\kappa}^{{\rm trans}(2)}\equiv 0$ and the transition process is the Airy$_{2\to 1}$, ${\cal A}_{2\to 1}$, discovered in~\cite{BFS07}:
\begin{equation}
{\cal A}_{2\to 1,0,\kappa}(\tau)\equiv{\cal A}_{2\to 1}(\tau).
\end{equation}
\end{remark}

\begin{proofOF}{Theorem~\ref{ThmTransition}}
To prove the result, we have to analyze the large-$T$ limit of the kernel in Proposition~\ref{PropStartingKernel} under the following scaling:
\begin{equation}
\begin{aligned}
t_i&=\frac{8\theta}{3}T-\tau_i (\pi'(\theta)+1)T^{2/3},\\
n_i&=M+\frac{2\theta}{3}T-\tau_i (\pi'(\theta)-1)T^{2/3},\\
x_i&=\tfrac12 t_i-2(n_i-M)-\widehat s_i T^{1/3},
\end{aligned}
\end{equation}
where
\begin{equation}
\widehat s_i=\left\{
\begin{array}{ll}
s_i,&\textrm{if }\tau_i\geq 0,\\
s_i-\tau_i^2 S_v S_h^{-2},&\textrm{if }\tau_i\leq 0.
\end{array}\right.
\end{equation}
Higher order in the development of $\pi(\theta-\tau_i T^{-1/3})$ are irrelevant since they corresponds to a $T^{-1/3}$ perturbation of $\pi'(\theta)$.

Then, we have to consider the rescaled and conjugated kernel
\begin{equation}
K^{\rm resc}_T(\tau_1,s_1;\tau_2,s_2):=\frac{{\rm Conj}_2}{{\rm Conj}_1}T^{1/3} K((n_1,t_1),x_1;(n_2,t_2),x_2)
\end{equation}
with ${\rm Conj}_i:=e^{t_i/2}(-1/2)^{n_i-M}(1/2)^{-(x_i+n_i-M)}$. We need to show that
\begin{equation}
\lim_{T\to\infty}K^{\rm resc}_T(\tau_1,s_1;\tau_2,s_2) = S_v^{-1}K^{{\rm trans}}_{M,\kappa S_v}(\tau_1/S_h,s_1/S_v;\tau_2/S_h,s_2/S_v).
\end{equation}

The first two terms of the kernel (\ref{KernelIntRepr}) are independent of $\alpha$ and their sum is the kernel without slow particles. This kernel was already analyzed in great detail in~\cite{BFS07} with the slight difference that the space-like setting introduced in~\cite{BF07} was not known yet. However, at the level of asymptotic analysis there are no relevant changes. Thus here we just indicate the key steps.

Let us first consider $\widehat K^{(1)}$ defined in (\ref{kernelK1}). After the change of variable $v=\tilde v-1$ it writes
\begin{equation}\label{eq6.17}
\frac{1}{(2\pi\I)^2}\int_{\Gamma_1}\dx \tilde v \oint_{\Gamma_{0,1-\tilde v}}\dx w \frac{2\tilde v-1}{w(w-\tilde v)(w+\tilde v-1)} \frac{e^{T f_0(w)+T^{2/3}f_{1,1}(w)+T^{1/3}f_{2,1}(w)}}{e^{T f_0(\tilde v)+T^{2/3}f_{1,2}(\tilde v)+T^{1/3}f_{2,2}(\tilde v)}}
\end{equation}
with
\begin{equation}
\begin{aligned}
f_0(w)&=\frac{8\theta}{3}\left(w+\tfrac14\ln((w-1)/w)\right), \\
f_{1,i}(w)&=-\tau_i(\pi'(\theta)+1)w-\tau_i(\pi'(\theta)-1)\ln(w-1)+\tfrac12(3-\pi'(\theta))\ln(w),\\
f_{2,i}(w)&=\widehat s_i \ln(w).
\end{aligned}
\end{equation}
From the analysis of Proposition~4 of~\cite{BFS07} we have that the steep descent paths in (\ref{eq6.17}) are chosen as illustrated in Figure~\ref{FigSteepTrans2}.
\begin{figure}
\begin{center}
\psfrag{1}[c]{$1$}
\psfrag{12}[c]{$\tfrac12$}
\psfrag{0}[c]{$0$}
\psfrag{a}[c]{$\alpha$}
\psfrag{(a)}[c]{(a)}
\psfrag{(b)}[c]{(b)}
\psfrag{v}[cb]{$\tilde v$}
\psfrag{z}[cb]{$\tilde z$}
\psfrag{w}[cb]{$w$}
\psfrag{q1}[cb]{$q$}
\psfrag{q2}[cb]{$q'$}
\psfrag{q3}[cb]{$q''$}
\psfrag{q4}[cb]{$q'''$}
\includegraphics[height=4cm]{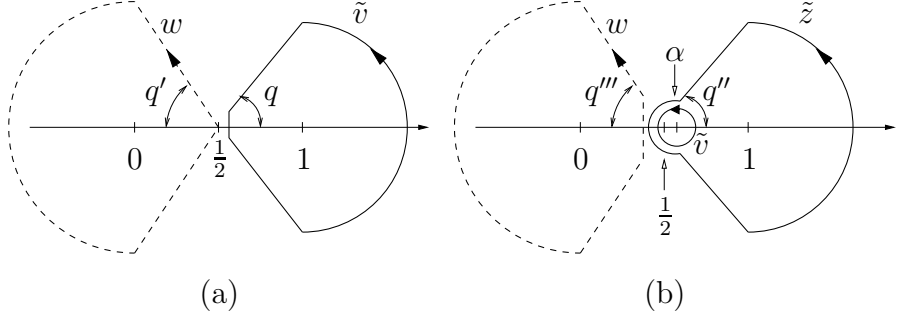}
\caption{Steep descents for (a) $\widehat K^{(1)}$ and (b) $\widehat K^{(2)}$. They satisfies $\pi/4<q<q'<\pi/2$, $\pi/4<q'',q'''<\pi/2$ and the local modifications around the critical point at $1/2$ are all only on the $T^{-1/3}$~scale.}
\label{FigSteepTrans2}
\end{center}
\end{figure}
Next, the Taylor expansion around the double critical point of $f_0(w)$ , which is at $w=1/2$, are given by
\begin{equation}
\begin{aligned}
f_0(w)&=f_0(1/2)-\frac{32\theta}{3}\frac{(w-1/2)^3}{3}+\Or((w-1/2)^4),\\
f_{1,i}(w)&=f_{1,i}(1/2)-\tau_i (5-3\pi'(\theta))(w-1/2)^2+\Or((w-1/2)^3),\\
f_{2,i}(w)&=f_{2,i}(1/2)+2\widehat s_i (w-1/2)+\Or((w-1/2)^2).
\end{aligned}
\end{equation}
The leading contribution to the kernel comes from the $T^{-1/3}$-neighborhood of the critical point. The conjugation terms are just the value of the exponential factor evaluated at the critical point. The $\Or(\cdots)$ term accounts into an error $\Or(T^{-1/3})$ smaller than the leading one. Then, after change of variable
\begin{equation}
w=\tfrac12+\tfrac12 w_1 T^{-1/3}/S_v,\quad \tilde v=\tfrac12+\tfrac12 w_2 T^{-1/3}/S_v,
\end{equation}
we get the final result
\begin{equation}
\begin{aligned}
&\lim_{T\to\infty}\frac{{\rm Conj}_2}{{\rm Conj}_1}T^{1/3} \widehat K^{(1)}_T((n_1,t_1),x_1;(n_2,t_2),x_2) \\
&=\frac{S_v^{-1}}{(2\pi\I)^2} \int_{\tilde\gamma_2}\dx w_2 \int_{\tilde \gamma_1}\dx w_1
 \frac{e^{w_2^3/3+\tau_2 w_2^2/ S_h-\widehat s_2w_2/S_v}}{e^{w_1^3/3+\tau_1 w_1^2/S_h-\widehat s_1w_1/S_v}}
 \frac{2w_2}{(w_1-w_2)(w_1+w_2)}.
\end{aligned}
\end{equation}

Consider now the $\alpha$-dependent term, $\widehat K^{(2)}$ defined in (\ref{kernelK2}). After the change of variable $v=\tilde v-1$ and $z=\tilde z-1$, (\ref{kernelK2}) becomes
\begin{multline}
\frac{1}{(2\pi\I)^3}\oint_{\Gamma_\alpha}\dx \tilde v \oint_{\Gamma_{1,\tilde v}}\dx \tilde z \oint_{\Gamma_{0,\alpha-\tilde v}}\dx w
\frac{(2\tilde z-1)(2\tilde v-\alpha)}{w(\tilde z-\tilde v)(\tilde z+\tilde v-1)(w-\tilde v)(w+\tilde v-\alpha)}\\
\times\left(\frac{w(w-\alpha)}{\tilde v(\tilde v-\alpha)}\right)^M \frac{e^{T f_0(w)+T^{2/3}f_{1,1}(w)+T^{1/3}f_{2,1}(w)}}{e^{T f_0(\tilde z)+T^{2/3}f_{1,2}(\tilde z)+T^{1/3}f_{2,2}(\tilde z)}}
\end{multline}
The leading term again comes from the $T^{-1/3}$-neighborhood of $1/2$. After the change of variables
\begin{equation}
w=\tfrac12+\tfrac12 w_1 T^{-1/3}/S_v,\quad \tilde z=\tfrac12+\tfrac12 w_2 T^{-1/3}/S_v,  \quad \tilde v=\tfrac12+\tfrac12 u T^{-1/3}/S_v
\end{equation}
and controlling the error terms as usual, we get
\begin{multline}
\lim_{T\to\infty}\frac{{\rm Conj}_2}{{\rm Conj}_1}T^{1/3} \widehat K^{(2)}_T((n_1,t_1),x_1;(n_2,t_2),x_2) =\frac{S_v^{-1}}{(2\pi\I)^3}
 \oint_{\Gamma_{\kappa}}\dx u \int_{\gamma_2}\dx w_2 \int_{\gamma_1}\dx w_1 \\
\frac{2w_2}{(w_2-u)(w_2+u)(w_1-u)}\left(\frac{w_1-S_v\kappa}{u-S_v\kappa}\right)^M \frac{e^{w_2^3/3+\tau_2 w_2^2/S_h-\widehat s_2w_2/S_v}}{e^{w_1^3/3+\tau_1 w_1^2/S_h-\widehat s_1w_1/S_v}}.
\end{multline}
Finally, concerning the integration paths, from the local structure around the critical point, see Figure~\ref{FigSteepTrans1}, we obtain the conditions illustrated in Figure~\ref{FigurePathTransition}.
\end{proofOF}

There is still one region where the $\alpha$ and $M$ dependence occurs in Figure~\ref{FigDiagram}. This is the transition between the Airy$_2$ process and ${\rm DBM}$. This is present for $\alpha\in (1/2,1)$ when $n\sim (1-\alpha)^2 t$, or in terms of $(\theta,\pi(\theta))$, it occurs for
\begin{equation}\label{eq5.25}
\pi(\theta)=\frac{2-2\alpha+\alpha^2}{\alpha(2-\alpha)}\theta.
\end{equation}
Consider the scaling (\ref{eq6.1}) with the condition (\ref{eq5.25}) and define the rescaled process as
\begin{equation}\label{eq5.27}
X_T(\tau)=\frac{x_n(t)-(t-2\sqrt{t(n-M)})}{-T^{1/3}}.
\end{equation}
In the large-$T$ limit $X_T$ converges to the following limit process.
\begin{defin}\label{DefTransitionGUE}
The process ${\cal A}_{{\rm DBM}\to 2}$ is the process with $m$-point distributions at $\tau_1<\tau_2<\ldots <\tau_m$ given by the Fredholm determinant,
\begin{equation}
 \Pb\left(\bigcap_{k=1}^m\left\{{\cal A}_{{\rm DBM}\to 2}(\tau_k)\leq s_k\right\} \right)
 = \det \left(\Id-\chi_s K_{{\cal A}_{{\rm DBM}\to 2}}\chi_s \right)_{L^2(\{\tau_1,\ldots,\tau_m\}\times \mathbb{R})}
\end{equation}
where $\chi_s(\tau_k,x)=\Id(x>s_k)$. The kernel is defined by
\begin{multline}\label{KernelGUEto2}
 K_{{\cal A}_{{\rm DBM}\to 2}}(\tau_1,s_1;\tau_2,s_2) =
-\frac{\exp\left(-\frac{((s_2-\tau_2^2)-(s_1-\tau_1^2))^2}{4(\tau_2-\tau_1)}\right)}{\sqrt{4\pi (\tau_2-\tau_1)}} \Id_{[\tau_2>\tau_1]}\\
+\frac{1}{(2\pi\I)^2}\int_{\gamma_2}\dx w_2 \int_{\gamma_1}\dx w_1
\frac{e^{w_2^3/3+\tau_2 w_2^2-(s_2-\tau_2^2)w_2}}{e^{w_1^3/3+\tau_1 w_2^2-(s_1-\tau_1^2)w_1}} \left(\frac{w_1}{w_2}\right)^M\frac{1}{w_1-w_2}
\end{multline}
Here $\gamma_2: e^{\pi\I/3} \infty\rightarrow e^{-\pi\I/3}\infty$, $\gamma_1: e^{-2\pi\I/3} \infty\rightarrow e^{2\pi\I/3}\infty$. Moreover, $\gamma_1,\gamma_2$ pass on the left of $0$ and they do not cross.
\end{defin}
With this definition, let us state the result.
\begin{thm}\label{ThmTransitionToGUE}
The process $X_T$ defined in (\ref{eq5.27}) converges to the process ${\cal A}_{{\rm DBM}\to 2}$, more precisely
\begin{equation}
\lim_{T\to\infty} X_T(\tau) = S_v {\cal A}_{{\rm DBM}\to 2}(\tau/S_h)
\end{equation}
in the sense of finite-dimensional distributions. The scaling coefficients $S_h$ and $S_v$ are given by
\begin{equation}
S_v=\left(\frac{2\theta\alpha}{(2-\alpha)(1-\alpha)}\right)^{1/3},\quad S_h=\frac{2 \alpha^{-1}}{1+\pi'(\theta)+\frac{1-\pi'(\theta)}{(1-\alpha)^2}} S_v^2.
\end{equation}
\end{thm}
\begin{proofOF}{Theorem~\ref{ThmTransitionToGUE}} The first part of the proof is in complete analogy to the one of Theorem~\ref{ThmTransition}, with the main difference being that the critical point is at $\alpha$ instead of $1/2$ (this explain why instead of $\kappa$ we get $0$).
We get the following expression (with $\tilde s_i=s_i-\tau_i^2$)
\begin{multline}
 -\frac{ \exp\left(-\frac{(\tilde s_2-\tilde s_1)^2}{4(\tau_2-\tau_1)}\right)}{\sqrt{4\pi (\tau_2-\tau_1)}}
 \Id_{[\tau_2>\tau_1]} + \frac{1}{(2\pi\I)^2} \int_{\gamma_2}\dx w_2 \int_{\gamma_1}\dx w_1
 \frac{e^{w_2^3/3+\tau_2 w_2^2-\tilde s_2w_2}}{e^{w_1^3/3+\tau_1 w_1^2-\tilde s_1w_1}}
 \frac{1}{w_1-w_2} \\
 + \frac{1}{(2\pi\I)^3}
 \oint_{\Gamma_{0}}\dx u \int_{\gamma_2}\dx w_2 \int_{\gamma_1}\dx w_1
 \frac{e^{w_2^3/3+\tau_2 w_2^2-\tilde s_2w_2}}{e^{w_1^3/3+\tau_1 w_1^2-\tilde s_1w_1}} \frac{1}{(w_2-u)(w_1-u)}\left(\frac{w_1}{u}\right)^M.
\end{multline}
Using the identities
\begin{equation}
\frac{1}{(w_2-u)(w_1-u)}=\frac{1}{w_1-w_2}\left(\frac{1}{w_2-u}-\frac{1}{w_1-u}\right)
\end{equation}
and
\begin{equation}
\frac{1}{w-u}=\frac{1}{w}\sum_{n\geq 0} (u/w)^n,\quad \textrm{if }|u|<|w|
\end{equation}
we compute the integral over $u$, which has a simple pole when $n=M-1$, letting to (\ref{KernelGUEto2}).
\end{proofOF}

A priori one might want to modulate the slow particle rate like in (\ref{eq5.2}) but around some $\alpha$ instead of $1/2$. This is however not a relevant change, since in a neighborhood of the curve in Figure~\ref{FigDiagram}, any point can be reached by fixing $\alpha$ and then choosing $\tau$ to get the desired value of $n/t$ or by choosing $\tau$ and then modulating $\alpha$.

Up to a change in the time direction, the kernel $K_{{\cal A}_{{\rm DBM}\to 2}}$ appeared in the context of sample covariance matrices~\cite{BBP04} (for $\tau_1=\tau_2=\tau$), and the extended version in TASEP with step initial conditions~\cite{SI07}, directed percolation with two set of parameters~\cite{BP07}, Brownian Motions with outliers~\cite{FAvM08}.

By looking at the diagram of Figure~\ref{FigDiagram} it is quite apparent that one should have the following limits:
\begin{equation}
\begin{aligned}
\lim_{\kappa\to\infty}{\cal A}_{2\to 1,M,\kappa}(\tau)&={\cal A}_{2\to 1}(\tau),\\
\lim_{v\to\infty}{\cal A}_{2\to 1,M,\kappa}(\tau+v)&=2^{1/3}{\cal A}_1(\tau/2^{2/3}),\\
\lim_{v\to\infty}\lim_{\kappa\to\infty}{\cal A}_{2\to 1,M,\kappa}(\tau-v)&={\cal A}_2(\tau),\\
\lim_{\kappa\to\infty} {\cal A}_{2\to 1,M,\kappa}(\tau-\kappa)&={\cal A}_{{\rm DBM}\to 2}(\tau),\\
\lim_{v\to\infty}{\cal A}_{{\rm DBM}\to 2}(\tau+v)&={\cal A}_{2}(\tau).
\end{aligned}
\end{equation}

\section{Regions where the slow particles do not matter}\label{SectNoSlow}
For completeness, we describe what happens in the region where the presence of slow particle is irrelevant. In the region where the density of particles is constant, the fluctuation of particles' positions are described asymptotically by the Airy$_1$ process. If the density of particles is decreasing (linearly in our case), then one has the Airy$_2$ process, and in the transition region where the density changes from constant to linearly decreasing, the process is the Airy$_{2\to 1}$ process. The computations are essentially the same as in~\cite{BFS07}, but easily extended to the setting of space-like paths. The only difference is that one has to control the new term coming from $\widehat K^{(2)}$.

Introduce the scaling on space-like paths described by a function $\pi(\theta)$ with $|\pi'|\leq 1$:
\begin{equation}\label{eq5.10}
\begin{aligned}
t(\tau,T) &= (\pi(\theta-\tau T^{-1/3})+\theta-\tau T^{-1/3})\, T,\\
n(\tau,T) &= M+[\pi(\theta-\tau T^{-1/3})-(\theta-\tau T^{-1/3})]\, T.
\end{aligned}
\end{equation}

\textbf{Case 1,} $n>\max\{\frac{1-\alpha}{2},1/4\}t$, i.e.\ $\pi(\theta)>\max\{\frac{3-\alpha}{1+\alpha},\frac53\}\theta$:
The rescaled process
\begin{equation}
X_T(\tau) = \frac{x_{n}(t)-(\frac12 t-2(n-M))}{-T^{1/3}}
\end{equation}
converges in the $T\to\infty$ limit to the Airy$_1$ process, ${\cal A}_1$,
\begin{equation}\label{Airy1}
 \lim_{T\to\infty}X_T(\tau) = S_v {\cal A}_1(\tau/S_h),
\end{equation}
where $S_v$ and $S_h$ are coefficients given by
\begin{equation}
 S_v = (\pi(\theta)+\theta)^{1/3},\quad S_h = \frac{4}{5-3\pi'(\theta)}S_v^2.
\end{equation}

\textbf{Case 2,} $\alpha\in (1/2,1]$ and $n\in ((1-\alpha)^2,1/4)t$, i.e.\ $\frac53\theta > \pi(\theta) >  \frac{2-2\alpha+\alpha^2}{\alpha(2-\alpha)}\theta$:
The rescaled process
\begin{equation}
X_T(\tau) =\frac{x_{n}(t)-(t-2\sqrt{t (n-M)})}{-T^{1/3}}
\end{equation}
converges in the large-$T$ limit to the Airy$_2$ process, ${\cal A}_2$,
\begin{equation}\label{Airy2}
 \lim_{T\to\infty}X_T(\tau) = S_v {\cal A}_2(\tau/S_h),
\end{equation}
where $S_v$ and $S_h$ are coefficients given by
\begin{equation}
\begin{aligned}
 S_v &= (\pi(\theta)+\theta)^{1/3}\left(\tfrac{\pi(\theta)-\theta}{\pi(\theta)+\theta}\right)^{-1/6} \left(1-\sqrt{\tfrac{\pi(\theta)-\theta}{\pi(\theta)+\theta}}\right)^{2/3},\\
  S_h &= \frac{2\left(1-\sqrt{\tfrac{\pi(\theta)-\theta}{\pi(\theta)+\theta}}\right)^{-1}}{(1-\pi'(\theta))\left(\frac{\pi(\theta)-\theta}{\pi(\theta)+\theta}\right)^{-1}+(1+\pi'(\theta))} S_v^2.
\end{aligned}
\end{equation}

\textbf{Case 3,} $\alpha\in (1/2,1]$ and $n\sim t/4$, i.e.\ $\pi(\theta)=\frac53 \theta$:
The rescaled process is given by
\begin{equation}
X_T(\tau) =
\left\{\begin{array}{ll}
{\displaystyle \frac{x_{n}(t)-(\frac12 t-2(n-M))}{-T^{1/3}}},&\textrm{if }n\geq t/4,\\[1em]
{\displaystyle \frac{x_{n}(t)-(t-2\sqrt{t(n-M)})}{-T^{1/3}}},&\textrm{if }n\leq t/4.
\end{array}\right.
\end{equation}
$X_T$ converges in the large-$T$ limit to the Airy$_{2\to 1}$ process, ${\cal A}_{2\to 1}$,
\begin{equation}\label{Airy21}
 \lim_{T\to\infty}X_T(\tau) = S_v {\cal A}_{2\to 1}(\tau/S_h),
\end{equation}
where $S_v$ and $S_h$ are coefficients given by
\begin{equation}
 S_v = (4\theta/3)^{1/3},\quad S_h = \frac{4}{5-3\pi'(\theta)}S_v^2.
\end{equation}

\section{Blocking wall regime}\label{SectWall}

In this section we study the case when $M=\infty$ and $\alpha=2$, so that the mean speed of the particles starting form $2\N$ (called $\alpha$-particles) is $1$, which is equal to the jump rate of particles starting from $2\Z_-$ (called \emph{normal particles}). We want to describe the large time behavior of a finite number of normal particles. For large time $t$, the $\alpha$-particles fluctuate on a $t^{1/3}$~scale, i.e., on a $t^{1/2}$~scale their behavior is essentially deterministic. $t^{1/2}$ is however the typical scale of fluctuations of the normal particles, which perform random walks except for being blocked by their right-neighbor. Thus, one expects that for large time, our system should be related to a set of non-intersecting Brownian motions with some particular condition at the origin (like absorption or reflection).

Theorem~\ref{thmWall} stated in Section~\ref{SectModel} is a direct consequence of the determinantal structure together with the following convergence of $K_\infty$ (defined in Corollary~\ref{CorMInfinity}) to the kernel $K^{{\rm aGUE}}$.
\begin{prop}\label{PropDiffusion}
Let
\begin{equation}\label{eq7.3}
t_i=\tau_i t,\quad x_i=t_i-\xi_i (2t_i)^{1/2}.
\end{equation}
Then,
\begin{equation}
\lim_{t\to\infty} \frac{A_1}{A_2}(2t_1)^{1/2} K_{\infty}((n_1,t_1),x_1;(n_2,t_2),x_2) = K^{{\rm aGUE}}((n_1,\theta_1),\xi_1;(n_2,\theta_2),\xi_2)
\end{equation}
where $\theta_i=\ln(\tau_i)$, with the conjugation factor $A_i=e^{-t_i}(t\tau_i/2)^{n_i/2}(-2)^{n_i}\tau_i^{1/2}$, and $K^{{\rm aGUE}}$ given in (\ref{aGUEmK}).
\end{prop}

Before proving the result, let us present an integral representation of the antisymmetric GUE minor kernel, since it is in that form that we obtain the result.
\begin{lem}\label{LemmaIntReprAsymGUE}
The antisymmetric GUE minor kernel has the following integral representation (after conjugation). Let $\tau_i:=e^{\theta_i}$ and $\e>0$. Then
\begin{equation}\label{eqIntGUEasym}
\begin{aligned}
&K^{{\rm aGUE}}((n_1,\theta_1),\xi_1;(n_2,\theta_2),\xi_2) \frac{B_2}{B_1}\\
=&
-\frac{2\sqrt{\tau_1}}{2\pi\I}\int_{\I\R+\e}\dx w e^{(\tau_1-\tau_2)w^2-2(\xi_1\sqrt{\tau_1}-\xi_2\sqrt{\tau_2})w}
\frac{1}{w^{n_2-n_1}}\Id_{[(n_1,\theta_1)\prec (n_2,\theta_2)]}\\
&- \frac{2\sqrt{\tau_1}}{2\pi\I}\int_{\I\R+\e}\dx w e^{(\tau_1-\tau_2)w^2-2(\xi_1\sqrt{\tau_1}+\xi_2\sqrt{\tau_2})w}
\frac{(-1)^{n_2+1}}{w^{n_2-n_1}}
\Id_{[(n_1,\theta_1)\prec (n_2,\theta_2)]}
\\
& + \frac{2\sqrt{\tau_1}}{(2\pi\I)^2}\oint_{\Gamma_0}\dx w_2 \int_{\I\R+\e}\dx w_1 \frac{e^{w_1^2\tau_1-2\xi_1\sqrt{\tau_1}w_1}}{e^{w_2^2\tau_2-2\xi_2\sqrt{\tau_2}w_2}}
\bigg(\frac{1}{w_1-w_2}+\frac{1}{w_1+w_2}\bigg)\frac{w_1^{n_1}}{w_2^{n_2}}
\end{aligned}
\end{equation}
with the paths non-crossing, i.e., $|w_2|<\e$, and $B_i=2^{n_i} e^{\theta_i (n_i+1)/2}$.
\end{lem}
\begin{proofOF}{Lemma~\ref{LemmaIntReprAsymGUE}}
We use the following two integral representations for the Hermite polynomials $H_n(x)$,
\begin{equation}
\begin{aligned}
H_n(x)&=\frac{2^n}{\I\sqrt{\pi}}e^{x^2}\int_{\I\R+\e}\dx w e^{w^2-2xw}w^n,\\
H_n(x)&=\frac{n!}{2\pi\I}\oint_{\Gamma_0}\dx z e^{-(z^2-2xz)} z^{-(n+1)},
\end{aligned}
\end{equation}
as well as the identities (with $0<q<1$) which can be found in~\cite{Jo04,KS96}
\begin{equation}\label{eq8.7}
\begin{aligned}
\frac{1}{\sqrt{\pi(1-q^2)}}\exp\left(-\frac{(x-qy)^2}{1-q^2}\right) &= e^{-x^2}\sum_{k=0}^\infty \frac{H_k(x)H_k(y) q^k}{\sqrt{\pi} 2^{k} k!},\\
\int_x^\infty \dx y e^{-y^2} H_n(y)&= e^{-x^2} H_{n-1}(x),\\
H_n(x)&=(-1)^n H_n(-x).
\end{aligned}
\end{equation}

Then, for $(n_1,\theta_1) \not\prec (n_2,\theta_2)$, we get (extending the sum to $\infty$ because the extra terms are identically zero) that $K^{{\rm aGUE}}((n_1,\theta_1),\xi_1;(n_2,\theta_2),\xi_2)$ is given by
\begin{equation}\label{eq8.8}
\frac{2^{n_1}}{2^{n_2}}\frac{4}{(2\pi\I)^2}\oint_{\Gamma_0}\dx z \int_{\I\R+\e}\dx w \frac{e^{w^2-2\xi_1 w}}{e^{z^2-2\xi_2 z}}\frac{w^{n_1+1}}{z^{n_2+2}}\sum_{\ell\geq 1}\left(\frac{z^2 e^{\theta_1}}{w^2 e^{\theta_2}}\right)^\ell.
\end{equation}
Now, by the change of variables $z=w_2 e^{\theta_2/2}=w_2\sqrt{\tau_2}$ and $w=w_1 e^{\theta_1/2}=w_1\sqrt{\tau_1}$ we obtain
\begin{equation}\label{eq8.9}
(\ref{eq8.8})=\frac{B_1}{B_2}\frac{4\sqrt{\tau_1}}{(2\pi\I)^2}\oint_{\Gamma_0}\dx w_2 \int_{\I\R+\e}\dx w_1 \frac{e^{w_1^2\tau_1-2\xi_1\sqrt{\tau_1} w_1}}{e^{w_2^2\tau_2-2\xi_2\sqrt{\tau_2}w_2}}\frac{w_1^{n_1+1}}{w_2^{n_2+2}}\sum_{\ell\geq 1}\left(\frac{w_2}{w_1}\right)^{2\ell}.
\end{equation}
Performing the sum over $\ell$,
\begin{equation}
\sum_{\ell\geq 1}\left(\frac{w_2}{w_1}\right)^{2\ell} = \frac{w_2^2}{w_1^2-w_2^2}\quad\textrm{for }|w_2|<|w_1|,
\end{equation}
and replacing in (\ref{eq8.9}) one obtains (\ref{eqIntGUEasym}).

Now consider $(n_1,\theta_1)\prec (n_2,\theta_2)$. Assume the following identity (proven below)
\begin{multline}\label{eq8.11}
F_{n_1}(\xi_1):=\frac{2^{n_1} \tau_1^{(n_1+1)/2}}{2^{n_2} \tau_2^{(n_2+1)/2}}\frac{2\sqrt{\tau_1}}{2\pi\I}\int_{\I\R+\e}\dx w e^{(\tau_1-\tau_2)w^2-2(\xi_1\sqrt{\tau_1}-\xi_2\sqrt{\tau_2})w}\frac{1}{w^{n_2-n_1}} \\
= \frac{2}{\sqrt{\pi}}e^{-\xi_1^2}\sum_{\ell=-\infty}^{n_2+1}\frac{e^{-(\theta_2-\theta_1)\ell/2}}{2^{n_2+1-\ell}(n_2+1-\ell)!} H_{n_1+1-\ell}(\xi_1) H_{n_2+1-\ell}(\xi_2).
\end{multline}
Then, the first two terms of (\ref{eqIntGUEasym}) are equal to
\begin{equation}
\frac{2}{\sqrt{\pi}}e^{-\xi_1^2}\sum_{\ell=-\infty}^{n_2+1}\frac{e^{-(\theta_2-\theta_1)\ell/2} H_{n_1+1-\ell}(\xi_1)}{2^{n_2+1-\ell}(n_2+1-\ell)!} \left(H_{n_2+1-\ell}(\xi_2)+(-1)^{n_2+1} H_{n_2+1-\ell}(-\xi_2)\right)
\end{equation}
and using the symmetry/antisymmetry properties of the Hermite polynomials, see (\ref{eq8.7}), we get a zero contribution for all odd $\ell$. From this follows the result. It remains to show (\ref{eq8.11}). We prove it by iteration, starting with $n_1=n_2$. The Gaussian integral gives
\begin{multline}\label{eq8.13}
\frac{2\sqrt{\tau_1}}{2\pi\I}\int_{\I\R+\e}\dx w e^{(\tau_1-\tau_2)w^2-2(\xi_1\sqrt{\tau_1}-\xi_2\sqrt{\tau_2})w} =\frac{\exp\left(-\frac{(\xi_1-\xi_2\sqrt{\tau_2/\tau_1})^2}{1-\tau_2/\tau_1}\right) }{\sqrt{\pi(1-\tau_2/\tau_1)}}\\
= 2\frac{e^{-\xi_1^2}}{\sqrt{\pi}}\sum_{\ell=-\infty}^{n_2+1} \frac{H_{n_2+1-\ell}(\xi_1)H_{n_2+1-\ell}(\xi_2) (\tau_2/\tau_1)^{{n_2+1-\ell}/2}}{2^{{n_2+1-\ell}} ({n_2+1-\ell})!}
\end{multline}
where we used (\ref{eq8.7}) with $q=\sqrt{\tau_2/\tau_1}$ and replaced $k$ by $n_2+1-\ell$. Next, notice that the function $F$ in (\ref{eq8.11}) satisfies
\begin{equation}
\int_x^\infty \dx y F_{n_1}(y) = F_{n_1-1}(x).
\end{equation}
Thus, to get $F_{n_2}$ we have to integrate $n_2-n_1$ times $F_{n_2}$. This is easily made using the integration formula in (\ref{eq8.7}) applied to $e^{-\xi_1^2}H_{n_2+1-\ell}(\xi_1)$ just $n_2-n_1$ times. This leads just to the shift in the index of the Hermite polynomials in (\ref{eq8.13}).
\end{proofOF}

\begin{proofOF}{Proposition~\ref{PropDiffusion}}
We prove that under the scaling (\ref{eq7.3})
\begin{equation}
\lim_{t\to\infty} \frac{C_1}{C_2}(2t_1)^{1/2} K_{\infty}((n_1,t_1),x_1;(n_2,t_2),x_2) = (\ref{eqIntGUEasym})
\end{equation}
where $\theta(\tau)=\ln(\tau)$, with the conjugation factor $C_i=e^{-t_i}(t/2)^{n_i/2}(-1)^{n_i}$. The kernel $K_\infty$ is given in Corollary~\ref{CorMInfinity}.

\begin{figure}[t]
\begin{center}
\psfrag{1}[cb]{$1$}
\psfrag{0}[cb]{$0$}
\psfrag{r}[cb]{$\rho$}
\psfrag{w}[cb]{$w$}
\psfrag{v}[cb]{$\tilde v$}
\psfrag{w2}[cc]{$\omega_2$}
\psfrag{case1}[cb]{(a)}
\psfrag{case23}[cb]{(b)}
\psfrag{case31}[cb]{(c)}
\psfrag{case32}[cb]{(d)}
\includegraphics[width=13.5cm]{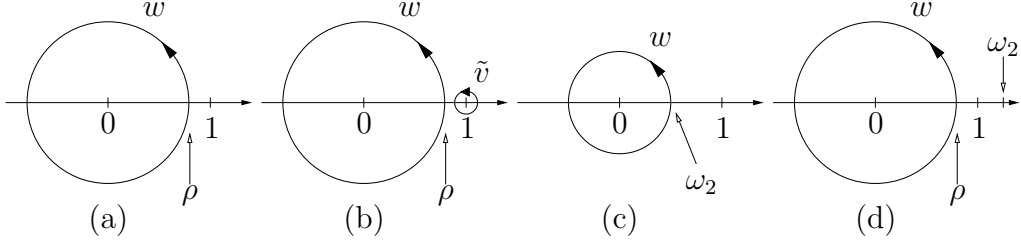}
\caption{Steep descent paths used for the different terms, with \mbox{$\rho=1-\e(t/2)^{-1/2}$}. (a) is for term (1); (b) for term (2) and (3a); (c) for term (3b) in case $\tau_1<\tau_2$; (d) for term (3b) in case $\tau_1>\tau_2$.}
\label{FigMinfinity}
\end{center}
\end{figure}

\vspace{1em}
\noindent\emph{(1) Term coming from $\hat\phi$.} We have
\begin{equation}\label{eq8.15}
-\hat\phi^{((n_1,t_1),(n_2,t_2))}(x_1,x_2) = -\frac{1}{2\pi\I}\oint_{\Gamma_0}\frac{\dx w}{w} \left(\frac{w-1}{w}\right)^{n_1-n_2} e^{t g_0(w)+t^{1/2} g_1(w)},
\end{equation}
with
\begin{equation}
\begin{aligned}
g_0(w)&=f_{0,1}(w)-f_{0,2}(w),\quad f_{0,i}(w)=\tau_i(w-\ln(w)),\\
g_1(w)&=f_{1,1}(w)-f_{1,2}(w),\quad f_{1,i}(w)=\xi_i\sqrt{2\tau_i}\ln(w).
\end{aligned}
\end{equation}
Consider $\tau_1>\tau_2$ (the case $\tau_1=\tau_2$ and $n_1<n_2$ is pretty easy). As steep descent path we can use $\Gamma_0=\{w=\rho e^{\I \phi}, \phi\in (-\pi,\pi]\}$ for any $\rho>0$, since $\Re(w-\ln(w))=\Re(w)-\ln(\rho)$. The critical point of $g_0$ is at $w=1$ and there the Taylor series are
\begin{equation}
\begin{aligned}
g_0(w)&=g_0(1)+\frac12(\tau_1-\tau_2)(w-1)^2+\Or((w-1)^3),\\
g_1(w)&=g_1(1)+(\xi_1\sqrt{\tau_1}-\xi_2\sqrt{\tau_2})\sqrt{2}(w-1)+\Or((w-1)^2).
\end{aligned}
\end{equation}
The leading contribution comes from the $t^{-1/2}$-neighborhood of $w=1$, which is however also a pole when $n_2>n_1$. Therefore we have to remain on its left and we choose $\rho=1-(2/t)^{1/2}\e$, $\e>0$. After controlling the error terms, we make the change of variable $w=1+t^{-1/2}\sqrt{2}W$ and take $W=\I\R-\e$ for any given $\e>0$. So, the leading contribution of (\ref{eq8.15}) is given by
\begin{equation}
\frac{(-1)^{n_1-n_2}C_2}{C_1}\sqrt{\frac{2}{t}}\frac{-1}{2\pi\I}\int_{\I\R-\e}\dx W W^{n_1-n_2} e^{(\tau_1-\tau_2)W^2+2(\xi_1\sqrt{\tau_1}-\xi_2\sqrt{\tau_2})W}.
\end{equation}
Finally, changing the variable $W=-w$ and multiplication by $C_1\sqrt{2t_1}/C_2$ leads to the first term in (\ref{eqIntGUEasym}).

\vspace{1em}
\noindent\emph{(2) Term coming from the second term in (\ref{eqCorMinfinity}).} After the change of variable $v=\tilde v-1$ we have
\begin{equation}\label{eq8.20}
\frac{1}{(2\pi\I)^2}\int_{\Gamma_1}\dx \tilde v\oint_{\Gamma_{0,1-\tilde v}}\dx w \frac{\left(\frac{w-1}{w}\right)^{n_1}}{\left(\frac{\tilde v-1}{\tilde v}\right)^{n_2}} \frac{e^{t f_{0,1}(w)+t^{1/2} f_{1,1}(w)}}{e^{t f_{0,2}(\tilde v)+t^{1/2} f_{1,2}(\tilde v)}}
\frac{2\tilde v-1}{w(\tilde v+w-1)(w-\tilde v)}
\end{equation}
The steep descent path for $w$ is chosen as above, while for $\tilde v$ we just take a circle around $1$ of radius smaller than $(2/t)^{1/2}\e$. With the variables called $\tilde v=1+t^{-1/2}\sqrt{2}V$ and $w=1+t^{-1/2}\sqrt{2}W$, the leading contribution of (\ref{eq8.20}) is
\begin{equation}
\frac{(-1)^{n_1-n_2}C_2}{C_1}\sqrt{\frac{2}{t}}\frac{1}{(2\pi\I)^2} \oint_{|V|<\e}\dx V \int_{\I\R-\e}\dx W \frac{e^{\tau_1 W^2+2\xi_1\sqrt{\tau_1}W}}{e^{\tau_2 V^2+2\xi_2\sqrt{\tau_2}V}} \frac{W^{n_1}}{V^{n_2}}\frac{1}{W-V}.
\end{equation}
Finally, the change of variable $W=-w_1$, $V=-w_2$ and multiplying by $C_1\sqrt{2t_1}/C_2$ leads to third term in (\ref{eqIntGUEasym}) (the part with $1/(w_1-w_2)$).

\vspace{1em}
\noindent\emph{(3) Terms coming from the third term in (\ref{eqCorMinfinity}).} Recall that we have to set $\alpha=2$ and after the change of variable $v=\tilde v-1$ we obtain
\begin{equation}\label{eq8.22}
\frac{1}{(2\pi\I)^2}\oint_{\Gamma_0}\dx w \oint_{\Gamma_{1,2-w}}\dx \tilde v \frac{e^{t_1 w}\left(\frac{w-1}{w}\right)^{n_1}}{w^{x_1}}\frac{\tilde v^{x_2}}{e^{t_2\tilde v}\left(\frac{\tilde v-1}{\tilde v}\right)^{n_2}}\frac{2\tilde v-1}{w(w+\tilde v-2)(1+\tilde v-w)}.
\end{equation}
\vspace{0.5em}
\emph{(3a) Term coming from the pole at $\tilde v=1$ of (\ref{eq8.22}).}
In this case the situation is almost identical as in case (2). The only difference is in the last factor, in particular, after the change of variables $\tilde v=1+t^{-1/2}\sqrt{2}V$ and \mbox{$w=1+t^{-1/2}\sqrt{2}W$} the last factor goes to $1/(W+V)$ instead of $1/(W-V)$. After the final change of variable is $W=-w_1$, $V=-w_2$ and multiplication by $C_1\sqrt{2t_1}/C_2$ leads to third term in (\ref{eqIntGUEasym}) (the part with $1/(w_1+w_2)$).\\[0.5em]
\emph{(3a ) Term coming from the pole at $\tilde v=2-w$ of (\ref{eq8.22}).} This term reads
\begin{equation}
\frac{1}{2\pi\I}\oint_{\Gamma_0}\dx w \frac{e^{t_1w+t_2(w-2)}(2-w)^{x_2}}{w^{x_1}}\frac{\left(\frac{w-1}{w}\right)^{n_1}}{\left(\frac{1-w}{2-w}\right)^{n_2}} =
\frac{1}{2\pi\I}\oint_{\Gamma_0}\dx w e^{t h_0(w)+t^{1/2} h_1(w)}\frac{\left(\frac{w-1}{w}\right)^{n_1}}{\left(\frac{1-w}{2-w}\right)^{n_2}},
\end{equation}
where
\begin{equation}
\begin{aligned}
h_0(w)&=(\tau_1+\tau_2)w+\tau_2(\ln(2-w)-2)-\tau_1\ln(w),\\
h_1(w)&=\sqrt{2\tau_1}\xi_1\ln(w)-\sqrt{2\tau_2}\xi_2\ln(2-w).
\end{aligned}
\end{equation}
There are two critical point of $h_0$, namely
\begin{equation}
\omega_1=1,\quad \omega_2=\frac{2\tau_1}{\tau_1+\tau_2},\textrm{ both in }[0,1].
\end{equation}
The steep descent path passes by the critical point the closest to the origin. For $\tau_1<\tau_2$, we have $\omega_2<1$ and the steep descent analysis gives readily a contribution of order
\begin{equation}
\frac{C_2}{C_1} e^{t h_0(\omega_2)-t h_0(1)}.
\end{equation}
It is easy to see that, with $\mu:=\tau_2/\tau_1$,
\begin{equation}
h_0(\omega_2)-h_0(1)=\tau_1 (1-\mu)(1+\ln(2)-\ln(\mu+1)) <0,\quad \textrm{for all }\mu>1.
\end{equation}
Thus in the $t\to\infty$ limit, the contribution goes to zero exponentially fast for $\tau_1<\tau_2$. Finally, consider $\tau_1>\tau_2$. Then, $1<\omega_2$. We choose the path as in case (1), but this time the Taylor series give
\begin{equation}
\begin{aligned}
h_0(w)&=h_0(1)+\frac{(\tau_2-\tau_1)}{2}(w-1)^2,\\
h_1(w)&=h_0(1)+(\xi_1\sqrt{\tau_1}+\xi_2\sqrt{\tau_2})\sqrt{2}(w-1)+\Or((w-1)^2).
\end{aligned}
\end{equation}
Also, we have a different sign in the prefactor and a factor $(-1)^{n_2}$ in the term $(w-1)^{n_1-n_2}$. This leads to the second term in (\ref{eqIntGUEasym}) and explains the differences with the first term of (\ref{eqIntGUEasym}), namely the $(-1)^{n_2+1}$ and the change $\xi_2\to -\xi_2$.
\end{proofOF}

\end{document}